\documentclass[
  aps,prx,
  onecolumn,          
  preprint,
  superscriptaddress,
  amsmath,amssymb,
  nofootinbib,
  floatfix
]{revtex4-2}

\usepackage{graphicx}
\usepackage[T1]{fontenc}
\usepackage[utf8]{inputenc}
\usepackage{braket}
\usepackage{bm}
\usepackage{color}
\usepackage{hyperref}
\usepackage{float}
\hypersetup{colorlinks=true,linkcolor=blue,citecolor=blue,urlcolor=blue}
\usepackage{xcolor}

\graphicspath{{figures/}}

\begin{document}

\title{Coherent Optical Control of a Single Donor Electron Spin}

\author{Yuxi Jiang}
\email{yxjiang@umd.edu}
\affiliation{Institute for Research in Electronics and Applied Physics and Joint Quantum Institute, University of Maryland, College Park, Maryland 20742, USA}
\affiliation{Department of Electrical and Computer Engineering, University of Maryland, College Park, Maryland 20742, USA}

\author{Amirehsan Alizadehherfati}
\affiliation{Institute for Research in Electronics and Applied Physics and Joint Quantum Institute, University of Maryland, College Park, Maryland 20742, USA}
\affiliation{Department of Electrical and Computer Engineering, University of Maryland, College Park, Maryland 20742, USA}

\author{Christine Falter}
\affiliation{JARA-Fundamentals of Future Information Technology, Forschungszentrum J\"ulich and RWTH Aachen University, 52062 Aachen, Germany}
\affiliation{Peter-Gr\"unberg-Institute (PGI-9), Forschungszentrum J\"ulich GmbH, 52428 J\"ulich, Germany}

\author{Jasvith R. Basani}
\affiliation{Institute for Research in Electronics and Applied Physics and Joint Quantum Institute, University of Maryland, College Park, Maryland 20742, USA}
\affiliation{Department of Electrical and Computer Engineering, University of Maryland, College Park, Maryland 20742, USA}

\author{Nils von den Driesch}
\affiliation{JARA-Fundamentals of Future Information Technology, Forschungszentrum J\"ulich and RWTH Aachen University, 52062 Aachen, Germany}
\affiliation{Peter-Gr\"unberg-Institute (PGI-10), Forschungszentrum J\"ulich GmbH, 52428 J\"ulich, Germany}

\author{Yurii Kutovyi}
\affiliation{JARA-Fundamentals of Future Information Technology, Forschungszentrum J\"ulich and RWTH Aachen University, 52062 Aachen, Germany}
\affiliation{Peter-Gr\"unberg-Institute (PGI-10), Forschungszentrum J\"ulich GmbH, 52428 J\"ulich, Germany}

\author{Jan Seelbinder}
\affiliation{JARA-Fundamentals of Future Information Technology, Forschungszentrum J\"ulich and RWTH Aachen University, 52062 Aachen, Germany}
\affiliation{Peter-Gr\"unberg-Institute (PGI-9), Forschungszentrum J\"ulich GmbH, 52428 J\"ulich, Germany}

\author{Alexander Pawlis}
\affiliation{JARA-Fundamentals of Future Information Technology, Forschungszentrum J\"ulich and RWTH Aachen University, 52062 Aachen, Germany}
\affiliation{Peter-Gr\"unberg-Institute (PGI-10), Forschungszentrum J\"ulich GmbH, 52428 J\"ulich, Germany}

\author{Edo Waks}
\email{edowaks@umd.edu}
\affiliation{Institute for Research in Electronics and Applied Physics and Joint Quantum Institute, University of Maryland, College Park, Maryland 20742, USA}
\affiliation{Department of Electrical and Computer Engineering, University of Maryland, College Park, Maryland 20742, USA}

\begin{abstract}

Coherent spin--photon interfaces for scalable quantum networks require bright optical transitions, long-lived spin coherence, and complete coherent spin control. Donor spins in direct-bandgap semiconductors combine strong donor-bound exciton transitions with a potentially nuclear-spin-free environment. However, coherent control of an individual optically active donor spin has remained elusive. Here, we demonstrate coherent optical control of a single Cl donor electron spin in ZnSe. Resonant excitation reveals spin relaxation dynamics and measures spin coherence through coherent population trapping (CPT). Off-resonant ultrafast pulses are implemented to drive spin Rabi oscillations and demonstrate Ramsey interference. Multiple dark resonances in the CPT spectrum further reveal hyperfine coupling between the electron and the \(I=3/2\) Cl nucleus. Together, these results demonstrate the first coherent optical control of a single donor electron spin in II--VI direct-bandgap semiconductors, establishing the essential components for exploiting optical donors in a scalable solid-state spin--photon platform.
\end{abstract}

\maketitle
\maketitle
\raggedbottom

\section{Introduction}

Solid-state spin qubits with optical interfaces can serve as stationary quantum memories and deterministic photon sources for distributed quantum information processing \cite{hCFl,CBxG,V4xE}. Realizing these capabilities in a single physical system requires a long-lived spin state that is coupled to efficient optical transitions at the individual-emitter level. While spins in several solid-state defect systems have shown extended coherence times reaching seconds \cite{tyryshkin2012electron}, the ability to connect distant spin qubits through efficient photonic channels remains challenging. Donor impurities in direct-bandgap semiconductors offer an alternate promising route to integrate spin-based quantum memory with efficient optical access \cite{kliX}. These donors can couple to donor-bound exciton (\(D^{0}X\)) states, exhibiting rapid radiative decay \cite{OM3Y,syNx,cODo} and strong coherent scattering under resonant excitation \cite{sanaka2009indistinguishable,sanaka2012entangling,jiang2026generation}. Moreover, II-VI semiconductors can provide a nuclear-spin-free environment through isotopic purification, potentially extending the electron spin coherence \cite{Tw7j,kirstein2021extended}. These properties make these donor spins compelling candidates for spin--photon interfaces.

The critical missing capability has been the coherent control of an individual donor-bound electron spin. Experiments on donor ensembles established several of the essential ingredients required for such control, including coherent population trapping \cite{Fu2005CPT} and ultrafast optical spin rotations \cite{kliX}. Extended spin coherence has been reported in ZnSe fluorine-donor ensembles \cite{HDFm,3ZEm}, and optical \cite{Linpeng2018Coherence,viitaniemi2022coherent,wang2026identification} and microwave \cite{FpcZ} control has been demonstrated in ZnO donor ensembles. Together, these studies established that the \(D^{0}X\) transitions can support both optical control and long-lived spin coherence across distinct direct-bandgap semiconductor hosts. At the single-donor level, optical spin initialization was demonstrated in fluorine impurities in ZnSe/ZnMgSe quantum wells \cite{Sleiter2013OpticalPumping,Kim2014Initialization}. However, to date, coherent control of a single donor electron spin remains experimentally elusive. Such control is essential for preparing arbitrary spin superposition states and generating spin–photon entanglement, providing a foundation for using individual donors as efficient quantum-network nodes.

Here, we demonstrate coherent control of a single donor-bound electron spin in ZnSe. We demonstrate optical spin initialization and readout, and measure a spin-relaxation time of \(T_{1}=18.5\pm3.8~\mu\mathrm{s}\), limited by charge tunneling rather than by an intrinsic spin-flip process. We then use detuned ultrafast optical pulses to coherently control the electron spin, observing spin Rabi oscillations and Ramsey interference. Finally, we drive two spin-selective optical transitions of the \(D^{0}X\) using resonant lasers and observe coherent population trapping. These measurements determine an inhomogeneous electron-spin dephasing time of \(T_{2}^{*}=21.9~\mathrm{ns}\). At higher coherent-driving powers, the dark resonance further separates into multiple features that reveal the hyperfine interaction between the electron and the donor nucleus, suggesting the potential for accessing the nuclear spin as an additional quantum memory. Together, these results establish an essential coherent quantum control framework for single optically active donors.

\section{Device and energy level structure}

\begin{figure}[htbp]
  \centering
  \includegraphics[width=0.85\linewidth]{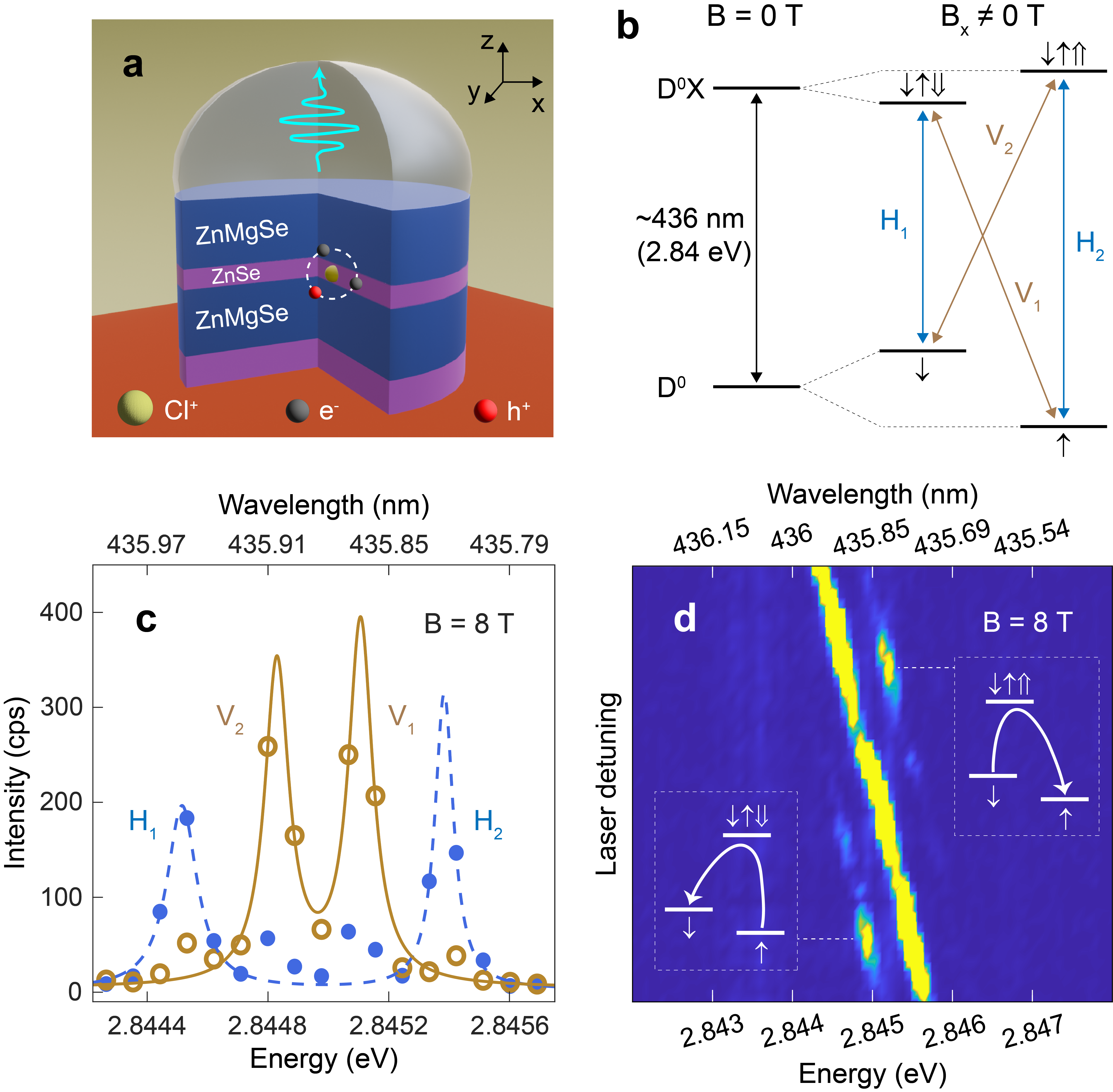}
  \caption{\textbf{A single optically addressable chlorine (Cl) donor in a ZnSe nanopost. a,} Schematic of the device structure, featuring a single Cl donor embedded in a \(\delta\)-doped ZnMgSe/ZnSe/ZnMgSe quantum well nanopost, topped with a dielectric solid immersion lens. \textbf{b,} Electronic level structure of the \(D^{0}X\) system. An external magnetic field in the Voigt geometry lifts the spin degeneracy. \(\uparrow\)/\(\downarrow\) represent the electron spin-up/spin-down states, and \(\Uparrow\)/\(\Downarrow\) represent the hole spin-up/spin-down states. \textbf{c,} Photoluminescence spectra of a single \(D^{0}X\) in an 8 T magnetic field. Yellow circles and blue solid dots denote emission in two orthogonal linear polarizations, respectively. Lines show Gaussian fits. \textbf{d,} Fluorescence spectrum obtained by scanning a tunable laser across the \(D^{0}X\) manifold. The excitation laser is \(H\)-polarized, while the signal is collected in the orthogonal \(V\)-polarization.}
  \label{fig1}
\end{figure}

The qubit studied here is an electron spin bound to an individual chlorine (Cl) donor in a ZnSe quantum well (Fig.~\ref{fig1}a). Single Cl donors are introduced and isolated in a \(\delta\)-doped layer in ZnSe with an ultralow density of \(10^{10}~\mathrm{cm}^{-2}\) using molecular-beam epitaxy \cite{OM3Y}. The doped ZnSe layer is confined between two ZnMgSe barriers, forming a quantum well that enhances the \(D^{0}X\) oscillator strength \cite{VhMY}. To improve photon-collection efficiency from a single emitter, the heterostructure is further patterned into nanoposts with directly integrated solid immersion lenses \cite{Gwnp}. Details of the device fabrication are provided in the Methods, and the material platform is described in greater detail in previous work \cite{Gwnp,OM3Y,jiang2026generation}.

The electronic level structure of the donor defect is shown in Fig.~\ref{fig1}b. The \(D^{0}\) state contains a single unpaired electron, whereas the \(D^{0}X\) state is formed by an electron singlet and a single heavy hole. When a magnetic field is applied in the Voigt geometry, these spin manifolds further split, giving rise to two sets of orthogonally linearly polarized optical transitions \cite{VhMY,0M3w}, providing the level structure required for optical control of the donor electron spin \cite{83Xt}.

We perform measurements on the ZnSe nanoposts at 3.6 K using a confocal optical microscopy system. We use nanoposts with a diameter of 200 nm, each of which statistically hosts 2-3 Cl atoms. To excite the sample and collect emission signals, we use a high-NA objective lens (NA = 0.68) in the confocal setup. The Methods section provides further details of the setup. By using a 405 nm laser to excite the quantum well above the ZnMgSe bandgap, we characterize the single \(D^{0}X\) emission with a characteristic second-order intensity correlation showing \(g^{(2)}(\tau=0)=0.02\) after background correction. See Supplementary Materials Section I for measurement results.

Fig.~\ref{fig1}c shows the \(D^{0}X\) photoluminescence spectrum at 8 T. The bound exciton exhibits four spectral peaks comprising two in \(H\)-polarization states and two in \(V\)-polarization states. By fitting the spectrum with Gaussian functions, we extract the donor electron \(g_{e} = 1.17\) and the \(D^{0}X\) heavy-hole \(3g_{\text{hh}} = 0.65\). The hole spin also exhibits an in-plane \(g\)-tensor anisotropy (see Supplementary Materials Section I) possibly due to reduced symmetry of the hole confinement potential \cite{0M3w}.

To address individual \(D^{0}X\) transitions for spin initialization, we scan a narrowband tunable laser across the spin manifold and detect the resulting fluorescence (Fig.~\ref{fig1}d). In this experiment, we use \(H\)-polarized excitation and collect \(V\)-polarized emission. To stabilize the donor charge state \cite{QTAb,jiang2026generation} and randomize its spin, we apply a weak above-band laser (at \(\sim 0.2\%\) of its saturation power). When the tunable laser is near resonance with the \(H_{1}\ (H_{2})\) transition, sidebands appear at frequency offsets of +132 GHz (-137 GHz) from the excitation frequency, corresponding to anti-Stokes (Stokes) spin-flip Raman scattering. The peak has a linewidth of 12 GHz (7.8 pm), approximately 11 times the lifetime-limited linewidth of \(\Gamma/(2\pi)=1.14~\mathrm{GHz}\), given that the excited-state decay rate is \(\Gamma=7.16~\mathrm{ns}^{-1}\) \cite{jiang2026generation,OM3Y}. This linewidth is \(\sim 60\%\) narrower than that reported in previous work \cite{jiang2026generation}. The narrower linewidth improves spectral selectivity at lower external magnetic fields, facilitating high-fidelity spin initialization and readout while potentially enabling microwave spin control together with optical detection \cite{Linpeng2018Coherence,FpcZ}. We further characterize the sideband emission rate in Supplementary Materials Section I.

\section{Spin relaxation dynamics}

\begin{figure}[H]
  \centering
  \includegraphics[width=0.85\linewidth]{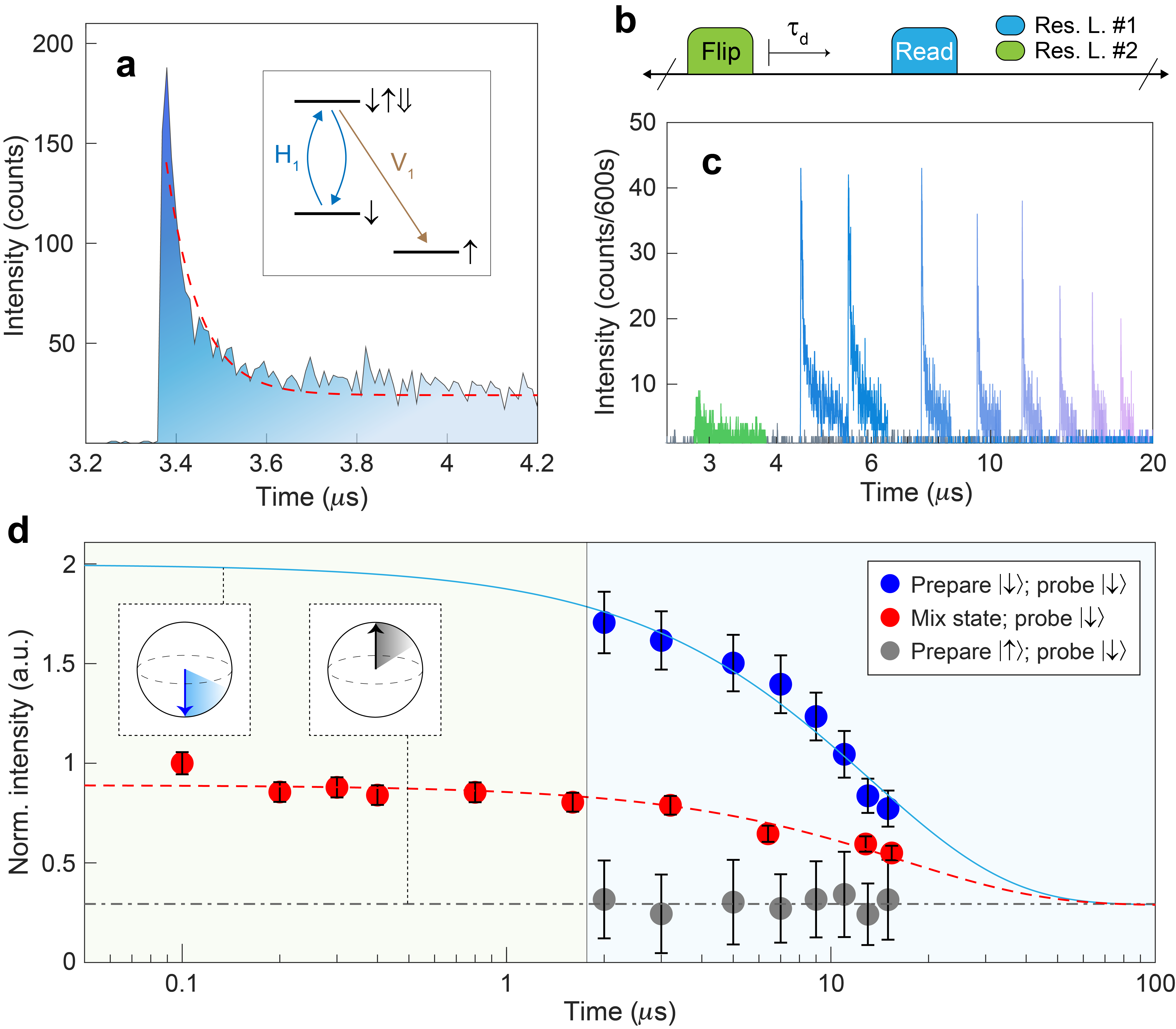}
  \caption{\textbf{Optical donor-spin initialization and measurement of spin relaxation (}\(T_{1}\)\textbf{). a,} Time-resolved fluorescence emission from the \(V_{1}\) transition of the \(D^0X\), under resonant excitation of its \(H_{1}\) transition. The inset illustrates the optical pumping scheme. \textbf{b,} Pump--delay--probe laser sequence for probing donor-spin relaxation dynamics. Delay time \(\tau_{d}\) is varied between measurement cycles to probe the spin population. \textbf{c,} Time-resolved fluorescence spectrum measured from the above pulse sequence. Eight traces are shown in different colors, corresponding to fluorescence emission at different \(\tau_{d}\). \textbf{d,} Normalized fluorescence intensity as a function of total evolution time. Measurements are performed under three initialization conditions: \textbf{(blue)} spin initialized to \(\lvert\downarrow \rangle\), \textbf{(grey)} spin initialized to \(\lvert\uparrow \rangle\), and \textbf{(red)} donor charged but spin not initialized. The \(\lvert\downarrow \rangle\) population is measured for all three conditions. Lines represent fits to the measurement data. Error bars represent photon-counting noise.}
  \label{fig2}
\end{figure}

To initialize the donor spin, we resonantly drive the \(H_{1}\) transition and measure fluorescence dynamics from the \(V_{1}\) transition (Fig.~\ref{fig2}a). Following a 0.5 \(\mu\)s above-band pulse that fully charges the donor and randomizes its spin, a 1 \(\mu\)s resonant pulse drives the \(H_{1}\) transition at 435.93 nm. The resulting fluorescence decays exponentially, indicating that the electron spin \(\lvert\downarrow \rangle\) population transfers through the spin-flip optical transition to the off-resonant \(\lvert\uparrow \rangle\) state, leading to spin polarization \cite{Sleiter2013OpticalPumping,Kim2014Initialization}. The spin-polarization rate is measured at different resonant powers (see Supplementary Materials Section II). These measurements calibrate driving field optical Rabi frequencies and the spin initialization fidelity.

We then use the pulse sequence shown in Fig.~\ref{fig2}b to measure the spin-relaxation time \(T_{1}\), providing an upper limit on the spin-coherence time. In the presence of donor discharging, we first initialize the electron spin to the \(\lvert\uparrow\rangle\) state using an above-band pulse followed by an \emph{Initialization} resonant laser pulse pumping the \(H_{1}\) transition. Then, a \emph{Flip} pulse drives the \(H_{2}\) transition to transfer the population back to the \(\lvert\downarrow\rangle\) state. We then let the donor evolve in the dark for a variable delay \(\tau_{d}\), after which we measure the \(\lvert\downarrow\rangle\) population using a \emph{Read} pulse that is again resonant with the \(H_{1}\) transition.

Figure~\ref{fig2}c shows fluorescence recorded during the \emph{Flip} and \emph{Read} pulses for several values of \(\tau_{d}\) using the above pulse sequence. The weak signal (green) observed during the \emph{Flip} pulse arises from the spectral filter being tuned to the \(V_{1}\) rather than the \(V_{2}\) transition, and does not indicate intrinsically weaker fluorescence (confirmed by additional measurements described in Supplementary Materials Section II). At the shortest delay, the fluorescence during the \emph{Read} pulse is \(1.71\) times stronger than that during the \emph{Initialization} pulse. This enhancement demonstrates efficient preparation of the spin state and excludes significant optically induced donor ionization, which would suppress the readout signal \cite{jiang2026generation}, as the emission from the \emph{Initialization} laser probes a donor-spin mixed state whereas the \emph{Read} pulse probes the \(\lvert\downarrow\rangle\) state. We also observe that when \(\tau_{d}\) increases, the fluorescence from the \emph{Read} pulse decreases, showing the loss of the prepared spin population during the dark interval.

We quantify the spin \(T_{1}\) from the time-resolved fluorescence intensity, as a direct measurement of the \(\ket{\downarrow}\) population after the interval \(\tau_{d}\). To separate the donor discharging and spin-flip processes, we perform the measurements under three spin-preparation conditions: the donor is initialized to the \(\ket{\downarrow}\) state or the \(\ket{\uparrow}\) state, or left in an unpolarized mixed state without spin initialization. For each condition, we normalize the integrated fluorescence from the \emph{Read} pulse as a ratio to the fluorescence from the pulse that initializes the spin.

Figure~\ref{fig2}d shows the intensity ratio as a function of \(\tau_{d}\) for these conditions. For a donor initialized in the \(\ket{\downarrow}\) state, the ratio is initially high and decreases as \(\tau_{d}\) increases. In contrast, for a donor initialized in the \(\ket{\uparrow}\) state, the ratio remains low, indicating that there is no measurable recovery of the \(\ket{\downarrow}\) population over \(\tau_{d}\). The mixed-state reference exhibits an intermediate signal intensity that likewise decreases with \(\tau_{d}\). The absence of the \(\ket{\downarrow}\) population recovery following the \(\ket{\uparrow}\) state preparation indicates that spin flips are not resolved within this measurement window. Instead, the observed decay of the \(\ket{\downarrow}\) population is primarily attributed to the donor charge tunneling. An exponential fit yields a charge-limited relaxation time of \(T_{1}=18.5 \pm 3.8~\mu\mathrm{s}\). The absence of population recovery further suggests that the intrinsic spin-flip time substantially exceeds the measured discharging-limited decay time. A one-sided confidence-interval analysis yields a \(90\%\) confidence lower bound of \(0.51~\mathrm{ms}\) for the spin-flip time (see Supplementary Materials Section II). Therefore, suppressing donor charge tunneling, for example through the DC electric field, could extend the current spin relaxation time toward its intrinsic spin-flip limit \cite{QTAb}.

\section{Coherent population trapping}

We next perform coherent population trapping (CPT) measurements on the \(D^{0}X\) optical transitions to investigate the donor-spin coherence. We select two spin ground states and a \(D^{0}X\)-excited state to form a \(\Lambda\) system, with the optical transitions resonantly driven by two laser fields \cite{Fu2005CPT,3uW4}. A stronger \emph{Control} laser drives the \(H_{1}\) transition, whereas a weaker \emph{Probe} laser is scanned across the \(V_{1}\) transition to measure its resonance fluorescence \cite{jiang2026generation}. The experimental configurations are described in Supplementary Materials Section III. Two-photon resonance occurs when the frequency difference between the two fields matches the spin ground-state splitting \cite{UqEG}. Under this condition, the two fields drive the system into a coherent spin-dark state, decoupled from optical excitation due to the formation of a superposition of the two spin states.

\begin{figure}[htbp]
  \centering
  \includegraphics[width=0.9\linewidth]{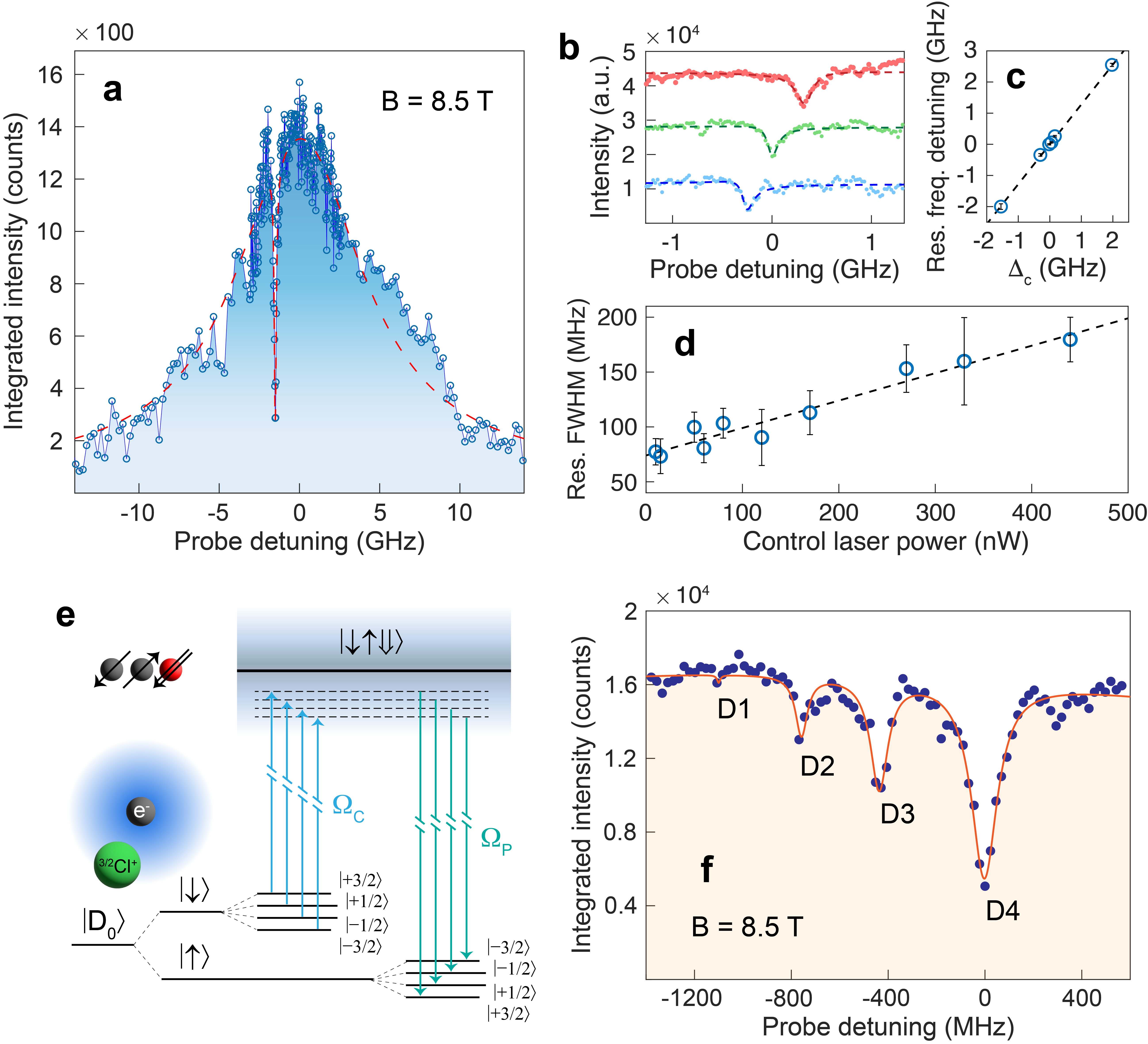}
  \caption{\textbf{Coherent population trapping (CPT) of a single donor electron spin. a,} Resonance fluorescence spectrum of the donor spin-\(\Lambda\) system under two-laser excitation. The \emph{Probe} laser is scanned across the \(V_{1}\) transition, while the resulting resonance fluorescence is measured. The \emph{Control} laser is resonant with the \(H_{1}\) transition (see Supplementary Materials Section III for details). \textbf{b,c,} CPT spectra for different \emph{Control} laser detunings, showing the corresponding shift of the two-photon resonance frequency. The dashed line shows a linear dependence of the resonance frequency on the laser detuning. \textbf{d,} CPT dip linewidth as a function of \emph{Control} laser power. The line shows a linear fit that describes power broadening. Error bars show fit uncertainty. \textbf{e,} Schematic of the hyperfine-resolved electron-nuclear-spin system. \textbf{f,} CPT spectrum under strong optical driving. Scanning the \emph{Probe} laser reveals four resonance dips. The curve shows a fit using four Lorentzian functions.}
  \label{fig3}
\end{figure}

Fig.~\ref{fig3}a shows the measured fluorescence intensity as a function of the \emph{Probe} laser detuning. A clear narrow dip appears at the two-photon resonance, centered within a broader fluorescence peak that reflects the inhomogeneously broadened \(D^0X\) optical transition \cite{jiang2026generation}. We model the CPT spectrum using a density-matrix approach, as described in Supplementary Materials Section III. Fitting the measured spectrum to the model yields a CPT dip linewidth of \(82~\mathrm{MHz}\) and a fluorescence peak linewidth of \(9~\mathrm{GHz}\).

To verify that the dip arises from coherent two-photon interference, we detune the \emph{Control} laser and repeat the \emph{Probe}-frequency scan (Fig.~\ref{fig3}b,c). For each scan, the \emph{Control} laser is fixed with a detuning of \(\Delta_{\rm C}\) from the \(H_{1}\) transition, while the \emph{Probe} laser is scanned across the \(V_{1}\) transition. As we vary \(\Delta_{\rm C}\), the dip position shifts accordingly (Fig.~\ref{fig3}b) with a linear dependence on \(\Delta_{\rm C}\) (Fig.~\ref{fig3}c). This confirms that the dark resonance follows the two-photon resonance condition rather than being a spectral feature fixed to the \(V_{1}\) transition.

We next investigate the CPT effect as a function of the \emph{Control} laser power. We scan the \emph{Probe} laser in frequency at a fixed power for different \emph{Control} powers and extract the dip linewidth from fits to each frequency scan. Fig.~\ref{fig3}d shows the measured dip linewidth as a function of \emph{Control} laser power. The linewidth increases with increasing power as a result of power broadening \cite{knappe2001characterization,javan2002narrowing}. Linearly extrapolating the measured linewidth to zero \emph{Control} power yields a linewidth of \(75~\mathrm{MHz}\), which provides a direct upper bound on the ground-state dephasing contribution to the CPT linewidth \cite{knappe2001characterization}.

We extract the donor spin dephasing time (\(T_{2}^{*}\)) from the CPT spectrum. We fit the spectrum in Fig.~\ref{fig3}a to the density-matrix model (Supplementary Materials Section III), from which we obtain the spin \(T_{2}^{*}=21.9~\mathrm{ns}\). At the two-photon resonance, the fluorescence is strongly suppressed with a dip contrast of \(0.83~\pm~0.02\), defined as the relative reduction in intensity from the peak maximum to the dip minimum. The significant emission suppression indicates the formation of a coherent spin-dark state. The density-matrix simulation yields a ground-state coherence of 0.31, showing that it generates 84\% of the ideal coherence, which is defined as \(\lvert\Omega_\mathrm{C}\Omega_\mathrm{P}^*\rvert/(\lvert\Omega_\mathrm{C}^2\rvert+\lvert\Omega_\mathrm{P}^2\rvert)=0.37\) in the absence of dephasing \cite{UqEG}.

To better understand the origin of the spin dephasing, we estimate the electron spin interaction with the nuclear spin bath in natural-abundance ZnSe. We use an analytical frozen-Overhauser model \cite{merkulov2002electron} and a discrete-lattice cluster-correlation-expansion calculation \cite{onizhuk2021pycce} to predict the spin dephasing time, both of which yield an estimated time constant of 23 ns (see Supplementary Materials Section IV). Quantum-well confinement may further reduce the \(T_2^*\) as a result of the electron localization. These estimates are in good agreement with the experimental value, supporting the role of the nuclear spin bath in limiting the electron spin coherence. Despite the use of natural ZnSe in the experiments, the dephasing time exceeds that of many electron-spin systems in nuclear-spin-rich materials, and can be further extended through isotopic purification \cite{kirstein2021extended}.

The Cl donor has a nuclear spin (I=3/2) that can couple to the donor electron through the Fermi-contact hyperfine interaction \cite{swift2020first,OM3Y}, producing four nuclear-spin sublevels and four possible CPT resonances (Fig.~\ref{fig3}e). The single dip observed in Fig.~\ref{fig3}a is attributed to optically induced nuclear polarization: the stronger \emph{Control} field polarizes the electron spin, which can polarize the Cl nucleus through hyperfine-mediated cross-relaxation \cite{jeffries1960dynamic,onur2016stabilizing}.

At stronger, comparable driving strengths of the two lasers (\(\Omega_{\mathrm{C}}=0.34\Gamma\) and \(\Omega_{\mathrm{P}}=0.32\Gamma\)), four adjacent dips emerge (D1--D4, Fig.~\ref{fig3}f), consistent with the expected hyperfine structure. Their unequal depths suggest unequal nuclear-spin populations \cite{wang2026identification}. We attribute the emergence of these additional resonances to off-resonant excitation by the two fields, followed by spontaneous decay that redistributes the nuclear-spin populations. This redistribution allows multiple CPT resonances to appear during the scan. Further discussion of the polarization mechanisms is provided in Supplementary Materials Section III. These observations reveal the coupled electron–nuclear spin structure of an individual Cl donor and suggest the potential for using its nuclear spin as an additional quantum memory controlled
through hyperfine coupling.

\section{Ultrafast optical spin rotation}

\begin{figure}[H]
  \centering
  \includegraphics[width=0.9\linewidth]{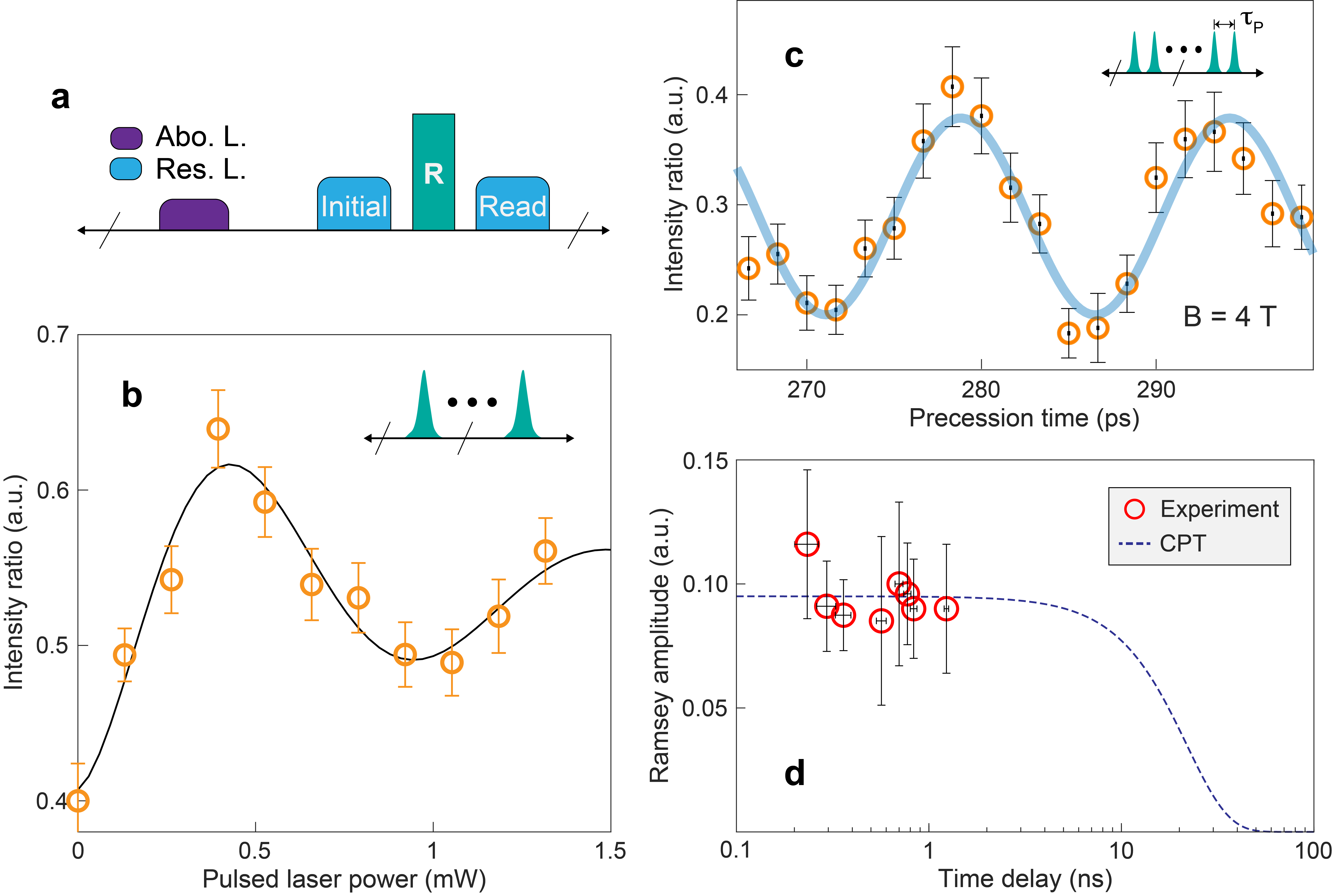}
  \caption{\textbf{Ultrafast coherent optical control of the donor spin.} \textbf{a,} Pulse sequence used for demonstrating the spin rotations. The spin is initialized and read out by resonant optical pumping, and rotated by ultrafast laser pulses. \textbf{b,} Fluorescence ratio as a function of the ultrafast laser power. \textbf{Inset}: sequence showing one ultrafast pulse used for each pumping cycle. \textbf{c,} Ramsey interference measured using two ultrafast pulses separated by a variable free-precession time \(\tau_{P}\). \textbf{Inset:} sequence showing two ultrafast pulses in each cycle, with a variable delay \(\tau_{P}\). \textbf{d,} Ramsey oscillation amplitude as a function of \(\tau_{P}\). Open circles represent amplitudes extracted from the experimental data, which are compared with the amplitude decay predicted using the \(T_{2}^{*}\) extracted from the CPT experiment. Error bars account for photon-counting noise and error propagation during normalization.}
  \label{fig4}
\end{figure}

Having established spin coherence of the \(D^0X\) through the CPT measurements, we next demonstrate coherent spin rotations using ultrafast laser pulses. To achieve this, we implement the pulse sequence shown in Fig.~\ref{fig4}a. The donor electron is first initialized in the \(\ket{\uparrow}\) state. It is then driven by an ultrafast pulse, which has a 1.6-ps pulse width and is frequency-doubled to 436.3 nm, \(700~\mathrm{GHz}\) red-detuned from the \(D^0X\) transitions to minimize unwanted excited-state population \cite{clark2007quantum,83Xt,kliX}. After a \(20~\mathrm{ns}\) delay from the first resonant pulse, we measure the \(\ket{\downarrow}\) population using a resonant laser. Further experimental details are provided in Supplementary Materials Section V.

Fig.~\ref{fig4}b shows the intensity ratio of the fluorescence from the \emph{Read} pulse to that from the \emph{Initialization} pulse. The ratio oscillates as a function of the ultrafast laser power. These oscillations provide direct evidence of the coherent Rabi oscillations of the donor spin \cite{83Xt,kliX,bqVK,CgfI}, whose angle is controlled by the ultrafast laser power. We model the spin rotation using a four-level density-matrix model, as detailed in Supplementary Materials Section V. Fitting the experimental data to the model yields a maximum rotation angle of 5\(\pi\)/2, limited by the available laser power. The oscillation becomes increasingly damped at higher powers, which we attribute to laser-induced dephasing \cite{Linpeng2018Coherence} and the excitation of free excitons \cite{kliX}.

Complete control of the donor spin requires rotation about a second axis, which we demonstrate using Larmor precession in a Ramsey pulse sequence (Fig.~\ref{fig4}c). We use a Ramsey interferometer with a mechanical delay stage to generate two ultrafast pulses with a variable interpulse delay \(\tau_{P}\). Each pulse has a fixed power of 0.2 mW, slightly below that required for a \(\pi/2\) rotation.

Fig.~\ref{fig4}c presents the measured intensity ratio as a function of \(\tau_{P}\), revealing Ramsey oscillations. These oscillations have an amplitude of \(0.12 \pm 0.03\) and a period of \(16 \pm 1\) ps, which agrees with the expected electron-spin precession period of 15.3 ps for \(g_{e} = 1.17\) at 4 T. The magnetic-field dependence of the oscillation period further confirms that the oscillation originates from the spin precession (see Supplementary Materials Section V).

To examine the spin coherence during free precession, we measure the Ramsey oscillation amplitude at various interpulse delays \(\tau_{P}\) of up to 1.2 ns, limited by the travel range of the mechanical delay stage. We extract the oscillation amplitude from each 30-ps delay window and assign it to the midpoint of the corresponding window.

Fig.~\ref{fig4}d plots the extracted oscillation amplitudes as a function of the delay. The measured amplitude exhibits no resolvable decay over 1.2 ns, indicating that the spin maintains its coherence during precession throughout the measurement window. Assuming a Gaussian envelope \(\exp\lbrack - (\tau/T_{2}^{*})^{2}\rbrack\) to describe the dephasing process \cite{Linpeng2018Coherence}, we use the previously measured \(T_{2}^{*}\) of 21.9 ns from the CPT experiments to predict an amplitude reduction of only 0.3\% at \(\tau_{P}=1.2\) ns, consistent with the Ramsey interference. While the interference currently exhibits moderate visibility limited by the excitation-induced dephasing, the Ramsey interference demonstrates coherent rotations of the donor-spin about two axes.

\section{Conclusion}

In conclusion, we report coherent optical control of a single donor electron spin in ZnSe. We demonstrate ultrafast spin control and coherent population trapping of the electron spin, and further resolve hyperfine-split CPT resonances arising from the interaction between the Cl nucleus and the bound electron. The donor spin exhibits a charge-tunneling-limited relaxation time of \(T_{1}=18.5\) \(\mu\)s and a dephasing time of \(T_{2}^{*}=21.9\) ns. Together, these results establish a coherently controlled donor spin--photon platform by coupling the spin state to efficient optical transitions. Looking ahead, improvements in epitaxial growth \cite{98d3} and active electric-field control \cite{QTAb} could mitigate donor charge tunneling, while isotopic purification of the host material could further extend electron-spin coherence \cite{Tw7j,kopteva2019spin}. Quantum frequency conversion \cite{J7a1} can further convert the emitted photons to telecommunication wavelengths for fiber-based quantum networks. These advances will unlock the potential for exploiting individual donors in scalable quantum photonic platforms.

\section{Methods}

\subsection{Fabrication process}

The sample was grown by molecular beam epitaxy, and the nanostructures were patterned using electron-beam lithography. The ZnSe/ZnMgSe quantum well structure was grown on GaAs (001) substrates. Before growth of the heterostructure, a 10-nm ZnSe buffer layer was deposited to ensure an abrupt interface between the III-V and II-VI semiconductors. A symmetric quantum well consisting of a 3.4-nm ZnSe layer sandwiched between two 27-nm ZnMgSe barriers was then grown. The ZnSe layer was doped with Cl at an ultralow density using a ZnCl\textsubscript{2} source. After growth, the structure was passivated in situ with a 10-nm aluminum oxide layer deposited by electron-beam evaporation. A SiO\textsubscript{x} solid immersion nanolens was fabricated on top of each nanopost using a negative electron-beam resist patterned by grayscale electron-beam lithography, followed by reactive-ion etching \cite{Gwnp}. Finally, the sample was cleaned using oxygen plasma to remove organic residues.

\subsection{Optical experimental setup}

We use a cryogenic confocal optical setup to excite the bound exciton and collect its emission. The sample is mounted on a three-dimensional piezoelectric stage in an Attodry 1000 closed-cycle Helium cryostat, with an objective lens of \(\mathrm{NA}=0.68\) mounted on the optical insert. The excitation beams are directed onto the sample at normal incidence by a mirror oriented at $45^\circ$ and positioned before the objective lens. An external magnetic field is applied in the sample plane, orthogonal to the laser propagation direction. Excitation lasers include a 405-nm single-mode laser diode (Thorlabs) for above-band excitation, two tunable diode lasers (435--443 nm; DL pro, TOPTICA) for resonant excitation, and a frequency-doubled Ti:Sapphire laser (400--450 nm; Mira 900) for ultrafast pulsed excitation. The two resonant excitation paths and the collection path each contain a half-wave plate, a quarter-wave plate, and a linear polarizer for polarization control. The collected reflection signal is coupled into a polarization-maintaining single-mode fiber and directed to a spectrometer (Princeton Instruments) and superconducting nanowire single-photon detectors (PhotonSpot) for spectral analysis and photon counting.

\bibliographystyle{apsrev4-2}
\bibliography{refs}

\section{Acknowledgements}

The authors thank Dr. Douglas L. Irving and Dr. Kelsey J. Mirrielees for valuable discussions.

\subsection{Funding}
The Waks group would like to acknowledge the support from the AFOSR (grant \#FA95502010250, grant
\#FA95502410266, and grant \#FA95502310667). The Pawlis
group would like to acknowledge the support from the
Deutsche Forschungsgemeinschaft (DFG, German Research Foundation) under Germany’s Excellence Strategy - Cluster of Excellence Matter and Light for Quantum Computing (ML4Q) EXC 2004/1-390534769.

\section{Author contributions}
Y.J. and E.W. conceived the study. Y.J., A.A., and C.F. performed the measurements. Y.J., A.A., and J.R.B. analyzed the data. C.F., N.v.d.D., Y.K., and J.S. fabricated the samples. A.P. and E.W. supervised the experimental work. All authors contributed to writing and revising the manuscript.

\section{Competing interests}
The authors declare that they have no competing interests.

\section{Data availability}
All data needed to evaluate the conclusions in the paper are present in the paper and/or the Supplementary Materials.

\end{document}


\maketitle

\setcounter{tocdepth}{1}
\tableofcontents

\clearpage

\section{Characterization of a single \(D^0X\) emission}
\subsection{Auto-correlation measurements}

\begin{figure}[htbp]
\renewcommand{\figurename}{FIG.}
\renewcommand{\thefigure}{S\arabic{figure}}
\centering
\includegraphics[width=\linewidth]{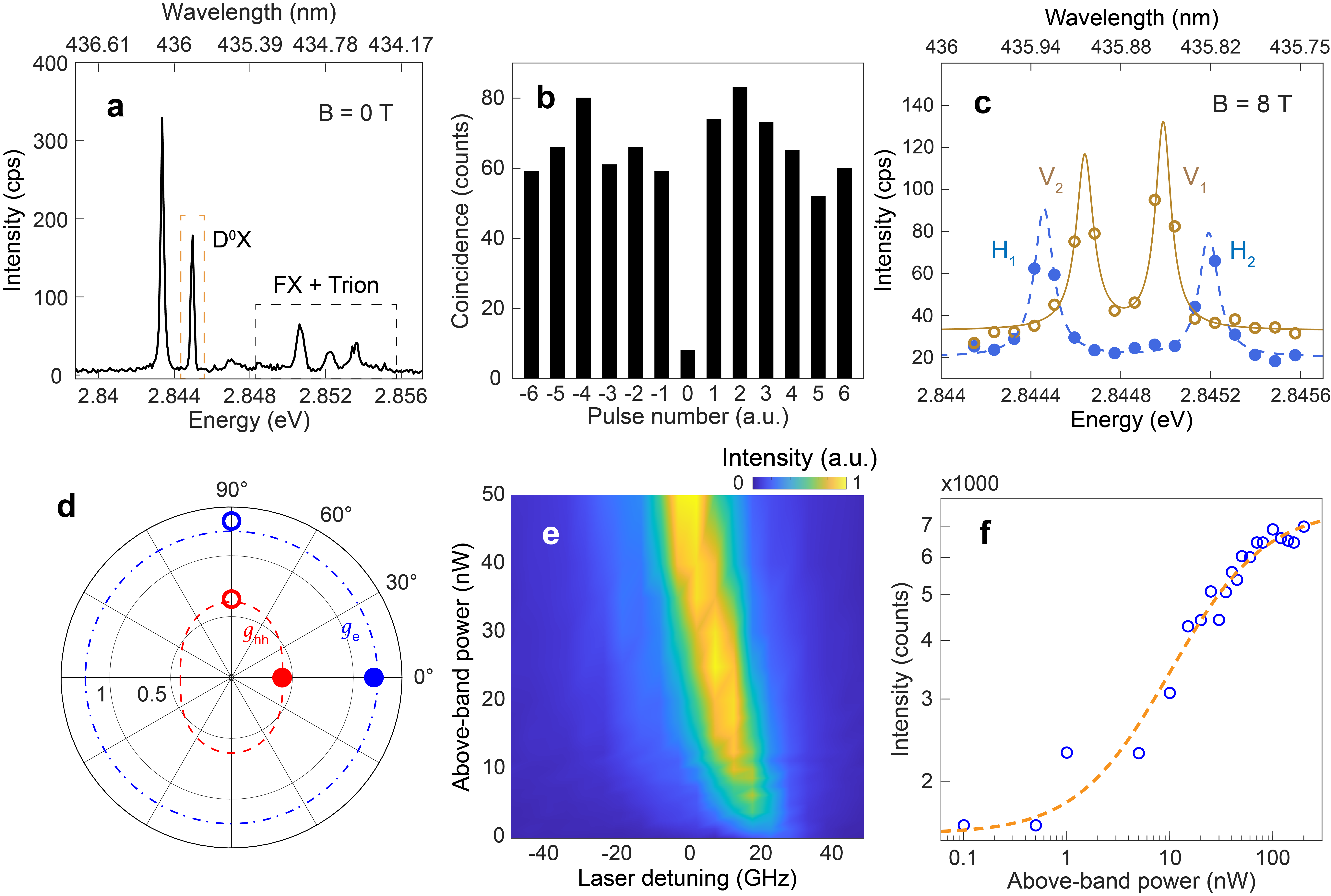}
\caption{\textbf{Single-photon emission and magneto-optical characterization of the \(D^0X\).}
\textbf{a,} Photoluminescence spectra of a single \(D^{0}X\) at 0 T magnetic fields. A single \(D^{0}X\) peak is red-detuned by several meV from the FX lines. \textbf{b,} Second-order intensity correlation measurement of the spectrally filtered \DzeroX{} photoluminescence. The zero-delay peak yields $g_{\rm raw}^{(2)}(0)=0.13$ before background subtraction. \textbf{c,} Polarization-resolved \DzeroX{} spectrum at $B=8$ T. Markers represent the measured data, and curves show Gaussian fits. \textbf{d,} Electron and heavy-hole $g$-factors extracted from magneto-photoluminescence spectra for two in-plane sample orientations. \textbf{e,} Raman sideband intensity as a function of resonant-laser detuning and above-band excitation power. \textbf{f,} Integrated $V_1$ Raman sideband intensity as a function of above-band excitation power. The dashed curve shows a fit to a saturation function \cite{pelton2002efficient}.}
\label{fig:S1}
\end{figure}

We first characterize the \(D^{0}X\) emission using 405 nm excitation above the ZnMgSe bandgap (Fig.~\ref{fig:S1}a). A narrow peak, labeled "\(D^{0}X\)" in the spectrum, corresponds to the bound exciton transition \cite{Gwnp}.

We measure the second-order intensity correlation function of \(D^0X\) fluorescence using a Hanbury--Brown--Twiss setup. A $76$-MHz pulse train with a pulse duration of 2.3 ps is used to excite the sample above the band gap. The \DzeroX{} emission is spectrally filtered using a tunable spectrometer filter with a 20-GHz bandwidth and then split by a non-polarizing $50{:}50$ fiber-based beam splitter. The photons are detected by two superconducting nanowire single-photon detectors. Photon arrival times are recorded using a PicoQuant MultiHarp 150 time-correlated single-photon counting module.

We integrate the area of each coincidence peak to extract the coincidence counts, which are plotted as a function of pulse number in Fig.~\ref{fig:S1}b. We observe photon antibunching, evidenced by the strongly suppressed zero-delay peak. The normalization baseline is obtained by averaging the areas of six peaks on each side of zero delay. Normalizing the zero-delay peak area to this baseline yields $g_{\rm raw}^{(2)}(0)=0.13$. We further correct for the background using the independently measured signal fraction, defined as $\rho=S/(S+B)$, where $S$ is the signal count rate and $B$ is the background count rate. Using $S=1.77\times10^4~\mathrm{s}^{-1}$ and $B=1.1\times10^3~\mathrm{s}^{-1}$, we obtain $\rho\approx0.94$. Assuming Poissonian background statistics and statistically independent signal and background, the corrected correlation is given by \cite{brouri2000photon}
\begin{equation}
g_{\rm corr}^{(2)}(0)
=\frac{g_{\rm raw}^{(2)}(0)-(1-\rho^2)}{\rho^2}.
\label{eq}
\end{equation}
Substituting the measured values yields $g_{\rm corr}^{(2)}(0)\approx0.02$.

\subsection{In-plane hole anisotropy}

We also measured the \DzeroX{} photoluminescence under above-band excitation after rotating the sample by $90^\circ$ in the sample plane, as shown in Fig.~\ref{fig:S1}c. At 8 T, the bound electron $g$-factor remains unchanged at $g_e=1.17$, whereas the heavy-hole $g$-factor changes from $3g_{hh}^{\parallel}=0.65$ to $3g_{hh}^{\perp}=0.41$. The orientation-dependent heavy-hole Zeeman splitting (Fig.~\ref{fig:S1}d), together with the unchanged electron $g$-factor, indicates an anisotropic in-plane magnetic response of the hole state, consistent with the heavy-hole $g$-factor anisotropy previously observed in ZnSe-based quantum wells \cite{0M3w,zhukov2021magnetic}.

\subsection{Effect of weak above-band excitation}

Weak above-band excitation stabilizes the donor charge state \cite{jiang2026generation,QTAb} and populates both electron-spin ground states through spontaneous decay of the bound exciton into either ground state. Fig.~\ref{fig:S1}e,f show an increase in Raman sideband intensity and a redshift of its center frequency with increasing above-band excitation power. With above-band excitation, the sideband intensity is enhanced by a factor of 4.85 relative to that measured without it, and the emission peak is red-shifted by approximately $20$~GHz. We attribute the enhancement to carrier-assisted recovery of the neutral donor charge state and the spectral shift to a change in the local electrostatic environment induced by above-band light \cite{QTAb}.

\clearpage

\section{Modeling of optical pumping and spin relaxation}

\subsection{Optical Rabi frequency and spin-polarization fidelity}

\begin{figure}[ht]
\renewcommand{\figurename}{FIG.}
\renewcommand{\thefigure}{S\arabic{figure}}
\setcounter{figure}{1}
  \centering
  \includegraphics[width=400px]{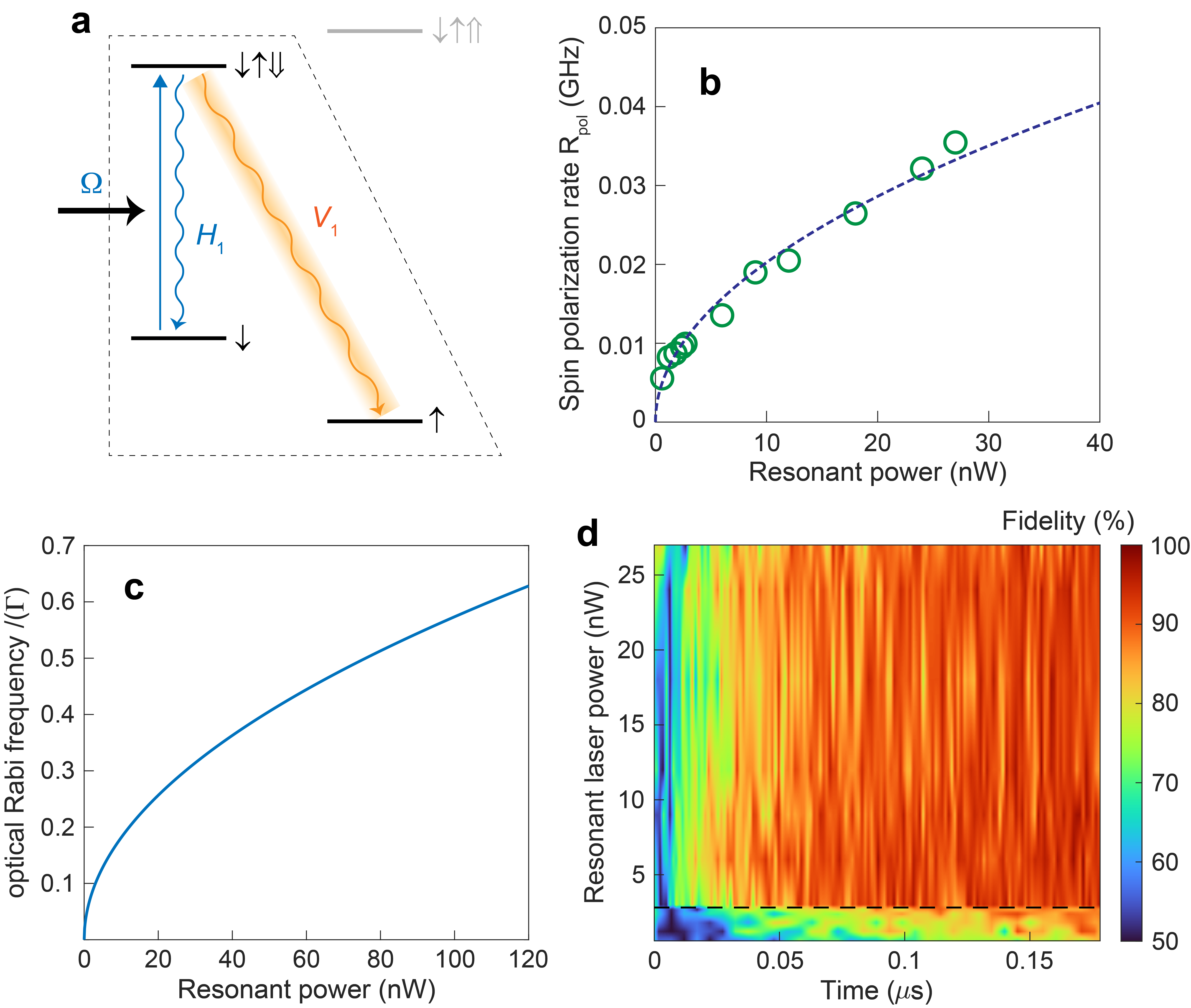}
  \caption{\textbf{Optical pumping mechanism, spin-polarization rate and fidelity, and optical Rabi frequency calibration.} 
  \textbf{a,} Schematic of resonant optical pumping for spin initialization.
  \textbf{b,} Extracted spin-polarization rate as a function of resonant excitation power. Blue circles show the measured rates, and the black dashed curve shows a fit to their power dependence. The corresponding Rabi frequency is shown on the right axis.
  \textbf{c,} Optical Rabi frequency of an on-resonant excitation laser as a function of its power.
  \textbf{d,} Spin-preparation fidelity as a function of resonant excitation power and pumping duration.}
  \label{fig:S2}
\end{figure}

Spin initialization is achieved through resonant optical pumping, as illustrated in Fig.~\ref{fig:S2}a. A laser resonantly driving the \(H_1\) transition couples the \(\ket{\downarrow}\) state to the excited state, which can decay through \(H_1\) to \(\ket{\downarrow}\) and \(V_1\) to \(\ket{\uparrow}\). Since the \(\ket{\uparrow}\) state is far-detuned from the coupling of the pumping laser, repeated excitation and decay progressively depletes the \(\ket{\downarrow}\) state and accumulates population in the \(\ket{\uparrow}\) state. The \(V_1\) fluorescence intensity tracks the \(\ket{\downarrow}\) population under fixed driving conditions. We therefore model the optical-pumping dynamics and compare to the measured time-dependent \(V_1\) fluorescence.

Let \(\Gamma\) denote the radiative decay rate of the excited state,
\begin{equation}
    \Gamma=\Gamma_{\rm sp}+\Gamma_{\rm sf},
\end{equation}
where \(\Gamma_{\rm sp}\) and \(\Gamma_{\rm sf}\) are the decay rates through the spin-preserving and spin-flipping transitions, respectively. The approximately equal intensities of the two transitions indicate a branching ratio
\begin{equation}
    \Lambda=\frac{\Gamma_{\rm sf}}{\Gamma_{\rm sp}}\simeq 1,
\end{equation}
and therefore
\begin{equation}
    \Gamma_{\rm sf}\simeq\Gamma_{\rm sp}\simeq\frac{\Gamma}{2}.
\end{equation}
Adapting the rate-equation treatment of weak-excitation optical pumping in Ref.~\cite{8Imr} to the approximately equal branching ratios of our system gives
\begin{equation}
    \Gamma_{\rm pol}=\frac{\Omega^2}{2\Gamma},
\end{equation}
where \(\Omega\) is the resonant laser Rabi frequency. To describe the full excitation-power dependence, including the saturation regime, we use
\begin{equation}
    \Gamma_{\rm pol}
    =\frac{\Gamma}{4}
    \frac{P/P_{\rm sat}}{1+P/P_{\rm sat}},
\end{equation}
where \(P\) is the resonant-excitation power and \(P_{\rm sat}\) is the saturation power. Assuming an initially unpolarized electron spin with no reverse pumping, the spin-initialization fidelity after a pumping duration \(\tau\) is
\begin{equation}
    F_{\rm init}(P,\tau)
    =1-\frac{1}{2}\exp\!\left[-\Gamma_{\rm pol}(P)\tau\right].
\end{equation}

The spin-polarization rate depends on the resonant excitation power. As shown in Fig.~\ref{fig:S2}b, the spin-polarization rate \(\Gamma_{\rm pol}\) is extracted by fitting the time-dependent fluorescence from the \(V_1\) transition at each excitation power to a single-exponential decay function. At low excitation powers, \(\Gamma_{\rm pol}\) increases approximately linearly with power, consistent with the weak-excitation relation \(\Gamma_{\rm pol}\propto P\). As the resonant excitation power increases, \(\Gamma_{\rm pol}\) approaches a rate of \(\Gamma/4\) and saturates, assuming equal spontaneous-emission branching ratios to the two spin ground states. Spectral wandering reduces the time-averaged resonant excitation efficiency. When the inhomogeneous linewidth greatly exceeds the natural linewidth, we account for this effect using an approximation to account for the effective power
\begin{equation}
P_{\rm eff}=P \frac{\Delta \nu_0}{\Delta \nu_{\rm inh}},
\end{equation}
where \(\Delta \nu_0=\Gamma/2\pi\) represents the natural linewidth of the bound exciton, and \(\Delta \nu_{\rm inh}\) is the inhomogeneously broadened bound exciton linewidth.

The optical Rabi frequency of the excitation laser is measured using a saturation power measurement \cite{jiang2026generation}, which yields a resonant excitation saturation power of 152 nW. Fig.~\ref{fig:S2}c shows the calibrated Rabi frequency of a resonant field as a function of the applied resonant excitation power.

We then extract the spin-preparation fidelity as a function of resonant excitation power and pumping duration from the time-resolved fluorescence, as shown in Fig.~\ref{fig:S2}d. The fidelity increases with both resonant excitation power and pumping duration, approaching a steady-state value of 0.95 at long pumping durations. Residual off-resonant excitation may repopulate the bright spin state due to the broadened linewidth and misalignment of the polarization setups, limiting the steady-state preparation fidelity.

\subsection{Intrinsic spin-flip time of the bound electron}

\begin{figure}[H]
    \renewcommand{\figurename}{FIG.}
    \renewcommand{\thefigure}{S\arabic{figure}}
    \setcounter{figure}{2}
    \centering
    \includegraphics[width=400px]{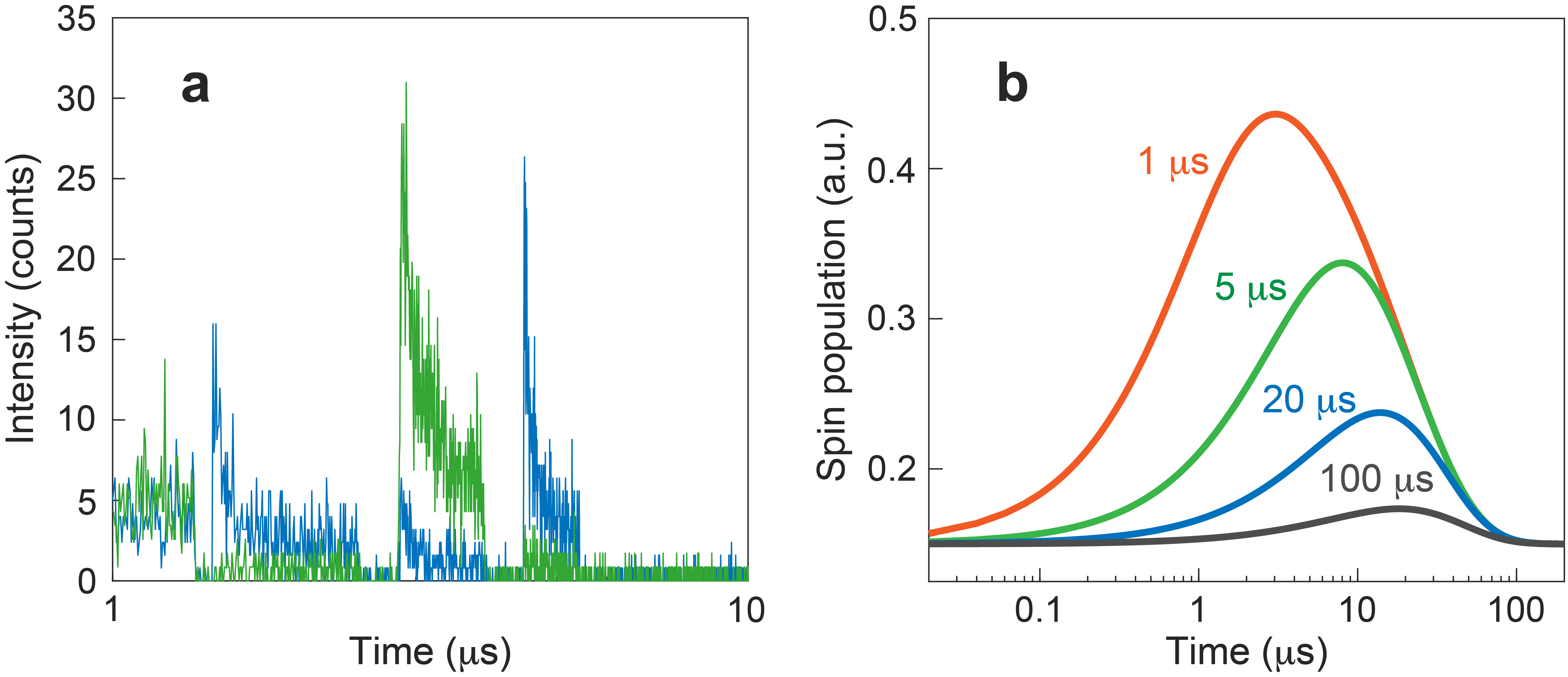}
    \caption{\textbf{Spin-population dynamics measured by time-resolved Raman fluorescence.}
    \textbf{a,} Time-resolved Raman fluorescence measured using the pump-delay-probe pulse sequence, with the spectral filter centered separately on the \(V_1\) (blue) and \(V_2\) (green) transitions.
    \textbf{b,} Calculated spin-population dynamics for intrinsic spin-flip time of \(T_{1,\mathrm{flip}}=1\), \(5\), \(20\), and \(100~\mu\mathrm{s}\), and an independently measured discharge time of \(T_e=18.5~\mu\mathrm{s}\).}
    \label{fig:S3}
\end{figure}

To verify that the weak fluorescence observed during the \emph{flip} pulse in Fig.~\ref{fig:S3}a does not arise from intrinsically weak emission on the relevant optical transition, we measure the time-resolved Raman fluorescence from \(V_1\) and \(V_2\) separately by tuning the spectral filter to each transition, using the pulse sequence shown in Fig.~2b. We observe a clear Raman fluorescence signal on the transition associated with the flip pulse when the corresponding spin ground state is populated.

During the variable delay \(\tau_d\) in Fig.~2b, the donor-electron population dynamics are governed by two competing processes: donor charge tunneling and intrinsic longitudinal spin relaxation. Charge tunneling ionizes the donor, rendering it optically dark and suppressing the fluorescence signal, whereas spin relaxation repopulates the initially depleted bright spin state, increasing the fluorescence intensity. Consequently, the bright-state population can initially rise even as the total neutral-donor population decreases, producing a transient fluorescence recovery followed by decay.

To constrain the intrinsic spin-flip time, we use a model for the fluorescence recovery
\begin{equation}
I(\tau;T_{1,\mathrm{flip}})
=
I_{\mathrm{0}}
+
A e^{-\tau/T_e}
\left[
1-e^{-\tau/T_{1,\mathrm{flip}}}
\right],
\label{eq:T1_recovery}
\end{equation}
where \(I_{\mathrm{0}}\) is the delay-independent background, \(A\) is the fluorescence amplitude associated with the bright spin state population, \(T_{e}\) is the donor charge tunneling time, and \(T_{1,\mathrm{flip}}\) is the intrinsic spin-flip time. The factor \(1-e^{-\tau/T_{1,\mathrm{flip}}}\) describes the repopulation of the bright state through spin-flip processes, while \(e^{-\tau/T_e}\) accounts for signal loss due to charge tunneling. Fig.~\ref{fig:S3}b shows the calculated recovery dynamics for several representative values of \(T_{1,\mathrm{flip}}\), illustrating how a shorter intrinsic spin-flip time produces a more pronounced transient fluorescence recovery before charge loss suppresses the signal.

In actual experiments, we did not observe clear spin-flip dynamics. To quantitatively constrain \(T_{1,\mathrm{flip}}\), we evaluate the profile statistic
\begin{equation}
\chi^2(T_{1,\mathrm{flip}})
=
\sum_i
\frac{
\left[
I_i-I(\tau_i;T_{1,\mathrm{flip}})
\right]^2
}{\sigma_i^2},
\label{eq:T1_chi2}
\end{equation}
where \(I_i\) and \(\sigma_i\) are the measured fluorescence and its standard uncertainty at delay \(\tau_i\), respectively. The measurement constrains \(T_{1,\mathrm{flip}}\) through a one-sided lower bound. We define the 90\% confidence boundary using
\begin{equation}
\Delta\chi^2
=
\chi^2(T_{1,\mathrm{flip}})
-
\chi^2_{\min}
=
\left[
\Phi^{-1}(0.90)
\right]^2
\simeq 1.64,
\end{equation}
where \(\Phi^{-1}\) is the inverse cumulative distribution function of the standard normal distribution. This analysis yields
\begin{equation}
T_{1,\mathrm{flip}}>0.51~\mathrm{ms}
\qquad
\text{(90\% confidence)}.
\end{equation}

\clearpage

\section{Coherent population trapping}

\subsection{Pulse sequence and polarization setup}

\begin{figure}[ht]
    \renewcommand{\figurename}{FIG.}
    \renewcommand{\thefigure}{S\arabic{figure}}
    \setcounter{figure}{3}
    \centering
    \includegraphics[width=400px]{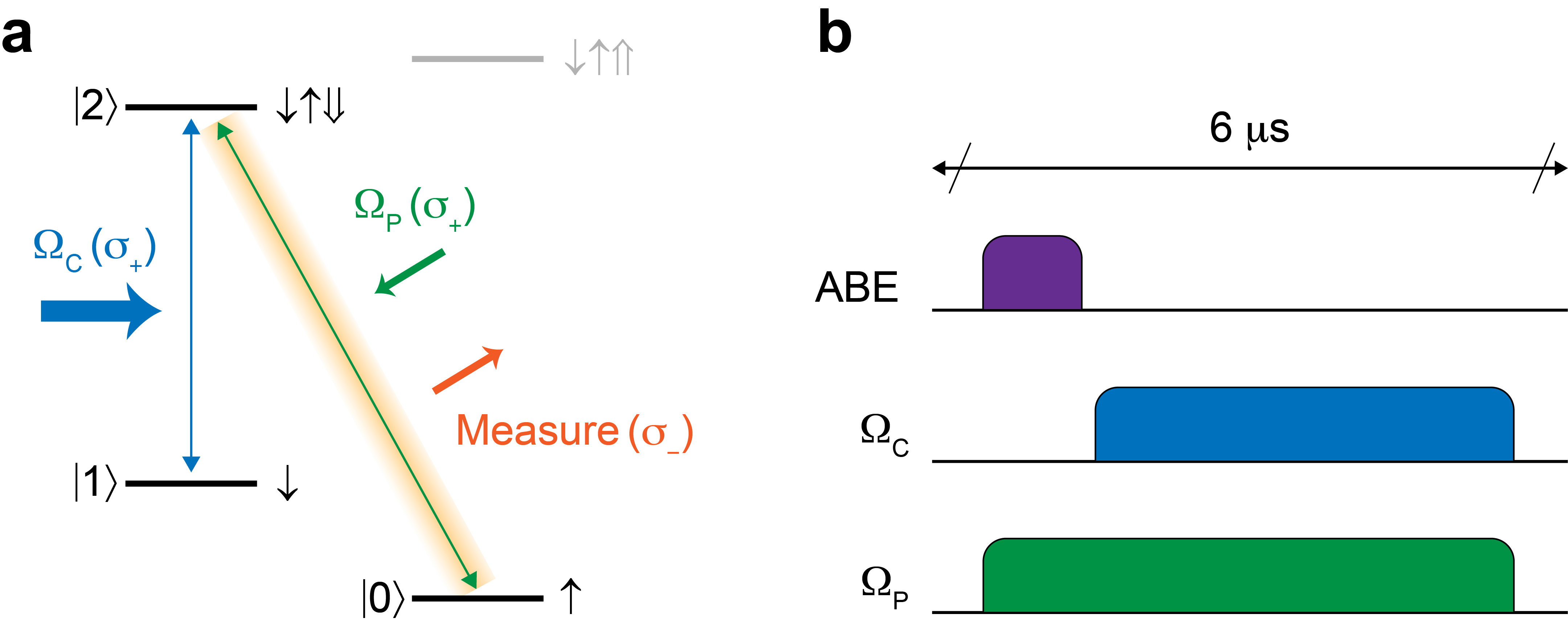}
    \caption{\textbf{CPT level scheme and pulse sequence.}
    \textbf{a,} $\Lambda$-type three-level system used for coherent population trapping.
    The \(Probe\) field $\Omega_{\mathrm P}$ couples the electron-spin ground state
    $\ket{0}$ to the common $D^{0}X$ excited state $\ket{2}$, while the \(Control\)
    field $\Omega_{\mathrm C}$ couples $\ket{1}$ to $\ket{2}$.
    Both excitation fields are prepared in the $\sigma_{+}$ polarization state,
    while resonance fluorescence is collected in the orthogonal $\sigma_{-}$ state.
    \textbf{b,} Pulse sequence used for the CPT measurement.
    A $1~\mu\mathrm{s}$ above-band excitation (ABE) pulse prepares the donor charge
    state, followed by a $5~\mu\mathrm{s}$ \(Control\) pulse.
    The \(Probe\) field is applied throughout the $6~\mu\mathrm{s}$ measurement cycle.}
    \label{fig:S4}
\end{figure}

The coherent population trapping (CPT) measurement is performed using the two electron-spin ground states, $\ket{0}$ and $\ket{1}$, coupled to the common $D^{0}X$ excited state $\ket{2}$ to form a $\Lambda$-type three-level system, as illustrated in Fig.~\ref{fig:S4}a. The \(Control\) laser is resonant with the $H_{1}$ transition, while the \(Probe\) laser is scanned across the $V_{1}$ transition. The pulse sequence used for the CPT measurement is shown in Fig.~\ref{fig:S4}b. Pulsed rather than continuous above-band excitation is used to reduce incoherent carrier generation and spin repopulation during the CPT measurement. The \(Control\) and \(Probe\) lasers are prepared in the same $\sigma_{+}$ polarization state with independent polarization control. Resonance fluorescence is collected in the orthogonal
$\sigma_{-}$ polarization state. This cross-circular excitation--collection
configuration suppresses residual scattering from the resonant laser while allowing detection of the resonance fluorescence near the
$V_{1}$ transition.

\subsection{Three-level density-matrix model}

We model the system as a three-level $\Lambda$ system comprising the
electron-spin ground states $\ket{0}$ and $\ket{1}$ and the donor-bound-exciton
state $\ket{2}$, following the density-matrix treatment of coherent population trapping and electromagnetically induced transparency
\cite{fleischhauer2005electromagnetically,UqEG}.

Under the rotating-wave approximation, the Hamiltonian is
\begin{equation}
    \frac{H}{\hbar}
    =
    -\Delta_{\mathrm P}\ket{2}\!\bra{2}
    -\delta\ket{1}\!\bra{1}
    +\frac{1}{2}
    \left(
        \Omega_{\mathrm P}\ket{2}\!\bra{0}
        +
        \Omega_{\mathrm C}\ket{2}\!\bra{1}
        +
        \mathrm{H.c.}
    \right),
    \label{eq:CPT_Hamiltonian}
\end{equation}
where $\Omega_{\mathrm P}$ and $\Omega_{\mathrm C}$ are the \(Probe\) and
\(Control\) Rabi frequencies, respectively. Their corresponding single-photon
detunings are $\Delta_{\mathrm P}$ and $\Delta_{\mathrm C}$, and
\begin{equation}
    \delta
    =
    \Delta_{\mathrm P}
    -
    \Delta_{\mathrm C}
\end{equation}
is the two-photon detuning.

The density matrix evolves according to
\begin{equation}
    \frac{d\rho}{dt}
    =
    -\frac{i}{\hbar}[H,\rho]
    +
    \mathcal{L}[\rho],
    \label{eq:CPT_master}
\end{equation}
where $\mathcal{L}$ accounts for spontaneous-emission and dephasing processes.
Using the Lindblad dissipator
\begin{equation}
    \mathcal{D}[L]\rho
    =
    L\rho L^\dagger
    -
    \frac{1}{2}
    \left\{
        L^\dagger L,\rho
    \right\},
\end{equation}
the spontaneous decay channel is written as
\begin{equation}
    \mathcal{L}_{\mathrm{rad}}[\rho]
    =
    \sum_{j=0,1}
    \mathcal{D}
    \left[
        \sqrt{\Gamma_{2j}}
        \ket{j}\!\bra{2}
    \right]\rho,
\end{equation}
where $\Gamma_{20}$ and $\Gamma_{21}$ are the decay rates from
$\ket{2}$ to $\ket{0}$ and $\ket{1}$, respectively. The total excited-state
population-decay rate is
\begin{equation}
    \Gamma
    =
    \Gamma_{20}
    +
    \Gamma_{21}.
\end{equation}

In the weak-\(Probe\) limit, $\Omega_{\mathrm P}\ll\Omega_{\mathrm C}$, the
zeroth-order population remains predominantly in the \(Probe\)-coupled ground
state, such that
\begin{equation}
    \rho_{00}\simeq1,
    \qquad
    \rho_{11},\rho_{22}\simeq0.
\end{equation}
Solving Eq.~\eqref{eq:CPT_master} in steady-state gives the excited-state population
\begin{equation}
    \rho_{22}
    =
    \frac{
        2\Omega_{\mathrm P}^{2}
        \left[
            \Omega_{\mathrm C}^{2}\gamma_{01}
            +
            4\gamma_{20}
            \left(
                \delta^{2}
                +
                \gamma_{01}^{2}
            \right)
        \right]
    }{
        \Gamma_{20}
        \left\{
            \left[
                4\Delta_{\mathrm P}\delta
                -
                \Omega_{\mathrm C}^{2}
                -
                4\gamma_{20}\gamma_{01}
            \right]^{2}
            +
            16
            \left(
                \Delta_{\mathrm P}\gamma_{01}
                +
                \delta\gamma_{20}
            \right)^{2}
        \right\}
    },
    \label{eq:CPT_population}
\end{equation}
where \(\gamma_{01}\) is the ground state dephasing rate. At the two-photon resonance, i.e., $\delta=0$, the two ground states form the dark-state
\begin{equation}
    \ket{D}
    =
    \frac{
        \Omega_{\mathrm C}\ket{0}
        -
        \Omega_{\mathrm P}\ket{1}
    }{
        \sqrt{
            |\Omega_{\mathrm C}|^{2}
            +
            |\Omega_{\mathrm P}|^{2}
        }
    },
    \label{eq:CPT_dark_state}
\end{equation}
which is decoupled from the transition to the excited state $\ket{2}$.

For the powers used in experiments, the weak-\(Probe\) approximation is not generally assumed. We therefore simulate the CPT spectra by numerically solving the three-level Lindblad master equation for the steady-state density matrix using \emph{QuTiP}. The fluorescence signal is modeled as a function of the steady-state excited state population
\begin{equation}
    I_{\mathrm{CPT}}(\delta)
    =
    A\rho_{22}^{\mathrm{ss}}(\delta)
    +
    I_{\mathrm{bg}},
    \label{eq:CPT_signal}
\end{equation}
where $A$ accounts for the radiative branching ratio together with the optical
collection and detection efficiencies, and $I_{\mathrm{bg}}$ is a detuning-independent background that accounts for the above-band excitation induced photoluminescence.

Slow spectral wandering broadens the one-photon optical resonance but does not affect the two-photon resonance linewidth, because the shift occurs at the optical bound-exciton state. To account for the observed optical
inhomogeneous broadening, we average the calculated steady-state fluorescence
over a Gaussian distribution of detunings,
\begin{equation}
    \overline{I}_{\mathrm{CPT}}
    \left(
        \Delta_{\mathrm P},\delta
    \right)
    =
    \int_{-\infty}^{\infty}
    d\Delta_0\,
    G
    \left(
        \Delta_0;\sigma_{\mathrm{opt}}
    \right)
    I_{\mathrm{CPT}}
    \left(
        \Delta_{\mathrm P}-\Delta_0,\delta
    \right),
    \label{eq:CPT_convolution}
\end{equation}
where
\begin{equation}
    G
    \left(
        \Delta_0;\sigma_{\mathrm{opt}}
    \right)
    =
    \frac{1}{
        \sqrt{2\pi}\sigma_{\mathrm{opt}}
    }
    \exp
    \left(
        -\frac{\Delta_0^2}{
            2\sigma_{\mathrm{opt}}^2
        }
    \right).
\end{equation}
Here, $\sigma_{\mathrm{opt}}$ denotes the spectral wandering of the optical state and is chosen to reproduce the approximately $9~\mathrm{GHz}$ peak envelope.

\subsection{Extraction of $T_2^*$ and CPT contrast}

The ground-state coherence time is extracted by fitting the measured CPT spectrum with the density-matrix model described above, including the spectral averaging of Eq.~\eqref{eq:CPT_convolution}. The optical decay rate and the \(Control\) and \(Probe\) Rabi frequencies are constrained by independent measurements. Based on these calibrations, the fit in the main text uses Rabi frequencies of \(\Omega_{\mathrm C}=0.29\Gamma\) and \(\Omega_{\mathrm P}=0.13\Gamma\), where \(\Gamma\) denotes the bound exciton decay rate.

The ground-state coherence can also be expressed in terms of the ideal CPT dark
state. From Eq.~\eqref{eq:CPT_dark_state}, the corresponding density operator
$\rho^{(D)}=\ket{D}\!\bra{D}$ has the off-diagonal element
\begin{equation}
    \left|
        \rho_{01}
    \right|
    =
    \frac{
        \left|
            \Omega_{\mathrm C}\Omega^*_{\mathrm P}
        \right|
    }{
        |\Omega_{\mathrm C}|^{2}
        +
        |\Omega_{\mathrm P}|^{2}
    }.
    \label{eq:CPT_coherence}
\end{equation}
This quantity represents the ground-state coherence of an ideal dark state for
the measured Rabi-frequency ratio and is used as a theoretical benchmark.

\subsection{Hyperfine-resolved CPT and nuclear-spin polarization}

To model the hyperfine-resolved CPT spectra, we extend the three-level model
to include the $I=3/2$ nuclear spin of the Cl donor. The Hamiltonian to describe this nuclear spin coupling is
\begin{equation}
\frac{H_{\mathrm{hf}}}{\hbar}
=
\omega_e S_z
-
\omega_n I_z
+
A_{\mathrm{hf}}
\left[
S_z I_z
+
\frac{1}{2}
\left(
S_+I_-
+
S_-I_+
\right)
\right],
\label{eq:hyperfine_Hamiltonian}
\end{equation}
where $\omega_e$ and $\omega_n$ are the electron and nuclear Zeeman frequencies, respectively, and $A_{\mathrm{hf}}$ is the isotropic electron--nuclear hyperfine coupling strength.

The $D^0$ ground-state manifold contains eight states corresponding to
$S=1/2$ and $I=3/2$. At the high magnetic fields used in our experiments, the electron Zeeman interaction dominates the hyperfine interaction, such that the eigenstates are approximated by the product states
$\lvert m_S,m_I\rangle$. The four nuclear-spin projections
$m_I=-3/2,-1/2,+1/2,+3/2$ shift the electron-spin splitting through the
secular hyperfine term $A_{\mathrm{hf}}S_zI_z$, giving rise to four
distinct two-photon resonance conditions. The much smaller nuclear Zeeman
splitting does not appreciably modify the experimentally resolved CPT
structure.

We construct a model comprising the eight hyperfine $D^0$ states. To first order, the optical transitions conserve the nuclear-spin projection, $\Delta m_I=0$, because the electric-dipole interaction does not directly couple to the nuclear spin angular momentum. Consequently, each $m_I$ manifold forms an approximately independent optical $\Lambda$ system, with its CPT resonance shifted by the corresponding hyperfine interaction.

The nonsecular hyperfine terms, $(S_+I_-+S_-I_+)/2$, couple states differing by
$\Delta m_S=\pm1$ and $\Delta m_I=\mp1$. Although this mixing is strongly
suppressed by the large electron Zeeman splitting, hyperfine-assisted
electron-spin relaxation and repeated optical pumping can produce a net
transfer of nuclear-spin population. In our experimental configuration,
optical pumping preferentially prepares the electron in the
$\lvert\uparrow\rangle$ state. For a relaxation pathway dominated by
$S_-I_+$ processes ($\Delta m_S=-1$ and $\Delta m_I=+1$) repeated pumping cycles progressively
transfer nuclear population toward larger $m_I$. The steady-state
population therefore accumulates predominantly in the
$m_I=+3/2$ manifold, producing a single dominant CPT resonance at low
optical power.

At stronger and more balanced optical excitation, off-resonant excitation increasingly addresses the other hyperfine manifolds. Repeated optical cycling redistributes population among the four $m_I$ states and competes with the electron-driven nuclear polarization. As a result, CPT resonances
associated with multiple nuclear-spin projections become visible. The hyperfine coupling $A_{\mathrm{hf}}$ primarily determines the spectral
separation of these resonances, whereas their relative amplitudes reflect the steady-state populations of the spin manifolds. This model accounts for the observed crossover from
a single CPT resonance at low optical power to four
hyperfine-resolved resonances at higher optical power.

\clearpage

\section{Numerical calculation of donor spin-dephasing time \texorpdfstring{$T_2^*$}{T2*}}

\subsection{Analytical model}

The electron bound to the Cl donor interacts with the isotopes of the ZnSe host crystal that have nonzero nuclear spin through hyperfine interactions. The Hamiltonian for this interaction is
\begin{equation}
    H_{\mathrm{hf}}
    =\sum_{j,k}x_{jk}A_{jk}\,\bm{S}\cdot\bm{I}_{jk},
\end{equation}
where $\bm{S}$ is the electron spin, $\bm{I}_{jk}$ is the nuclear spin from species $j\in\{\mathrm{Zn},\mathrm{Se}\}$ at sublattice site $k$, $A_{jk}$ is the hyperfine interaction parameter for that specific nuclear spin, and $x_{jk}$ is a binary isotope-occupation variable (1 if the site contains a spinful isotope and 0 otherwise). For a magnetic field along $z$ that is sufficiently large to suppress electron--nuclear spin flip-flops, the secular part is \cite{merkulov2002electron,de2003theory}
\begin{equation}
    H_{\mathrm{hf}}^{\mathrm{sec}}=S_z h_z,
    \qquad
    h_z=\sum_{j,k}x_{jk}A_{jk}I_{z,jk}.
\end{equation}

The quantity $h_z$ has units of energy and is the longitudinal Overhauser shift of the electron-spin splitting. Nuclear-spin dynamics are much slower than the nanosecond free-precession interval considered here. Thus, $h_z$ is effectively constant during one Ramsey sequence, while repeated measurements sample different nuclear configurations. For a fixed value of $h_z$, the relative phase accumulated by the two electron-spin states during a free-evolution time $t$ is $h_zt/\hbar$. Averaging this phase over the nuclear ensemble gives the normalized transverse coherence
\begin{equation}
    L_R(t)
    =\left\langle\exp\!\left(-\frac{i h_z t}{\hbar}\right)\right\rangle
    =\int dh_z\,P(h_z)\exp\!\left(-\frac{i h_z t}{\hbar}\right),
\end{equation}
where $P(h_z)$ is the probability distribution of the quasi-static Overhauser shift over different nuclear-bath configurations. The Ramsey signal contains a controlled oscillatory phase multiplied by $L_R(t)$; therefore, the magnitude of $L_R(t)$ is the Ramsey-fringe envelope. In other words, the free-induction decay is the Fourier transform of the static Overhauser distribution.

Because the electron overlaps with many host nuclei, the distribution of $h_z$ is approximately Gaussian. Its characteristic function is therefore
\begin{equation}
    L_R(t)=\exp\!\left(-\frac{\sigma_h^2t^2}{2\hbar^2}\right).
\end{equation}
Using the Ramsey fit convention
\begin{equation}
    L_R(t)=\exp\!\left[-\left(\frac{t}{T_{2}^*}\right)^2\right],
\end{equation}
we obtain
\begin{equation}
    T_{2}^*=\frac{\sqrt{2}\hbar}{\sigma_h}.
\end{equation}
The goal is to extract the standard deviation of $h_z$, thereby determining the dephasing coherence time.

At the experimental temperature, the nuclear Zeeman energies are much smaller than $k_BT$. We therefore treat the host nuclei as unpolarized and uncorrelated. For a nucleus of spin $I_j$,
\begin{equation}
    \left\langle I_{z,jk}\right\rangle=0,
    \qquad
    \left\langle I_{z,jk}^2\right\rangle
    =\frac{I_j(I_j+1)}{3}.
\end{equation}
Using the statistical independence of different nuclei and isotope occupations, the mean Overhauser shift vanishes and its variance is
\begin{equation}
    \sigma_h^2
    \equiv\left\langle h_z^2\right\rangle
    =\sum_{j,k}\left\langle x_{jk}\right\rangle A_{jk}^2
    \left\langle I_{z,jk}^2\right\rangle
    =\sum_{j,k}y_j A_{jk}^2\frac{I_j(I_j+1)}{3},
\end{equation}
where $y_j$ is the natural isotope abundance. We use the envelope-function approximation \cite{de2003theory} to find $A_{jk}$. We write the donor-electron wavefunction as
\begin{equation}
    \Psi(\bm{r})=F(\bm{r})u_c(\bm{r}),
\end{equation}
where $u_c(\bm{r})$ is the periodic conduction-band Bloch function and $F(\bm{r})$ is the slowly varying donor envelope, normalized according to
\begin{equation}
    \int |F(\bm{r})|^2\,d^3r=1.
\end{equation}
The contact coupling to a particular nucleus of species $j$ at position $\bm{R}_{jk}$ is written as \cite{merkulov2002electron,syperek2011long}
\begin{equation}
    A_{jk}=A_j^{\mathrm{HF}}v_0\left|F(\bm{R}_{jk})\right|^2,
\end{equation}
where $A_j^{\mathrm{HF}}$ is the material hyperfine constant for species $j$, independent of position. The quantity $v_0=a_0^3/4$ is the zincblende primitive-cell volume, which contains one Zn site and one Se site. Therefore, the density of either sublattice is $1/v_0$. Substituting the site-dependent coupling gives
\begin{equation}
    \sigma_h^2
    =\sum_j y_j\frac{I_j(I_j+1)}{3}
    \left(A_j^{\mathrm{HF}}\right)^2v_0^2
    \sum_k\left|F(\bm{R}_{jk})\right|^4.
\end{equation}
The envelope changes slowly on the lattice scale, so the sum over either sublattice may be replaced by
\begin{equation}
    \sum_k\left|F(\bm{R}_{jk})\right|^4
    \simeq\frac{1}{v_0}\int|F(\bm{r})|^4\,d^3r.
\end{equation}
The host-nuclear variance is therefore
\begin{equation}
    \sigma_h^2
    =v_0\int|F(\bm{r})|^4\,d^3r
    \sum_j y_j\frac{I_j(I_j+1)}{3}
    \left(A_j^{\mathrm{HF}}\right)^2.
\end{equation}
This form is useful because the actual electron wavefunction enters only through its inverse participation volume,
\begin{equation}
    J=\int|F(\bm{r})|^4\,d^3r.
\end{equation}

We begin with an electron bound to a donor in a bulk ZnSe crystal. The wavefunction can be modeled as a nominal isotropic hydrogenic envelope,
\begin{equation}
    F(r)=\frac{1}{\sqrt{\pi a^{*3}}}\exp\!\left(-\frac{r}{a^*}\right).
\end{equation}
One finds
\begin{equation}
    J_{\mathrm{bulk}}
    =\int|F(\bm{r})|^4\,d^3r
    =\frac{1}{8\pi a^{*3}},
\end{equation}
and hence
\begin{equation}
    \sigma_h^2
    =\frac{v_0}{8\pi a^{*3}}
    \sum_j y_j\frac{I_j(I_j+1)}{3}
    \left(A_j^{\mathrm{HF}}\right)^2.
\end{equation}

The calculation below includes only the $^{77}\mathrm{Se}$ and $^{67}\mathrm{Zn}$ host nuclei. The central Cl nucleus is excluded from this calculation because the present analysis is intended to estimate dephasing from the surrounding Zn and Se host-nuclear-spin bath. The parameters used in the host-bath estimate are listed below.

\begin{table}[H]
    \centering
    \begin{tabular}{@{}p{0.42\linewidth}p{0.30\linewidth}p{0.16\linewidth}@{}}
        \toprule
        \textbf{Quantity} & \textbf{Value} & \textbf{Source} \\
        \midrule
        ZnSe lattice constant, $a_0$
        & $0.566$~nm
        & Ref.~\cite{HDFm} \\

        Primitive-cell volume, $v_0=a_0^3/4$
        & $0.0453$~nm$^3$
        & \\

        $^{77}\mathrm{Se}$ abundance, spin
        & $y_{\mathrm{Se}}=0.0758$, $I_{\mathrm{Se}}=1/2$
        & Refs.~\cite{syperek2011long,HDFm} \\

        $^{77}\mathrm{Se}$ hyperfine constant
        & $A_{\mathrm{Se}}=33.6~\mu\mathrm{eV}$
        & Refs.~\cite{syperek2011long,HDFm} \\

        $^{67}\mathrm{Zn}$ abundance, spin
        & $y_{\mathrm{Zn}}=0.0411$, $I_{\mathrm{Zn}}=5/2$
        & Ref.~\cite{HDFm} \\

        $^{67}\mathrm{Zn}$ hyperfine constant
        & $A_{\mathrm{Zn}}=3.7~\mu\mathrm{eV}$
        & Ref.~\cite{HDFm} \\
        \bottomrule
    \end{tabular}
\end{table}

The nominal radius follows from the hydrogenic relation
\begin{equation}
    a^*=\frac{\hbar}{\sqrt{2m^*E_b}},
\end{equation}
using $m^*=0.16m_0$, $\epsilon=8.66$, and a representative Cl-donor binding energy near $26.9$~meV as an estimate.\textsuperscript{13} For an extracted radius of $a^*=3.0$~nm, the isotope contributions are
\begin{equation}
    \sigma_{\mathrm{Se}}^2=1.46\times10^{-3}~\mu\mathrm{eV}^2,
    \qquad
    \sigma_{\mathrm{Zn}}^2=1.12\times10^{-4}~\mu\mathrm{eV}^2.
\end{equation}
Thus,
\begin{equation}
    \sigma_h
    =\sqrt{\sigma_{\mathrm{Se}}^2+\sigma_{\mathrm{Zn}}^2}
    =39.7~\mathrm{neV}.
\end{equation}
This corresponds to a frequency standard deviation
\begin{equation}
    \sigma_f=\frac{\sigma_h}{2\pi\hbar}=9.6~\mathrm{MHz},
\end{equation}
or a one-dimensional rms Overhauser field of approximately $0.62$~mT for $g=1.1$. The predicted Gaussian Ramsey time is
\begin{equation}
    T_{2,\rm G}^*=\frac{\sqrt{2}\hbar}{\sigma_h}=23.4~\mathrm{ns}.
\end{equation}
The Se isotope contributes approximately $93\%$ of the total variance.

\begin{figure}[H]
    \renewcommand{\figurename}{FIG.}
    \renewcommand{\thefigure}{S\arabic{figure}}
    \setcounter{figure}{4}
    \centering
    \includegraphics[width=\linewidth]{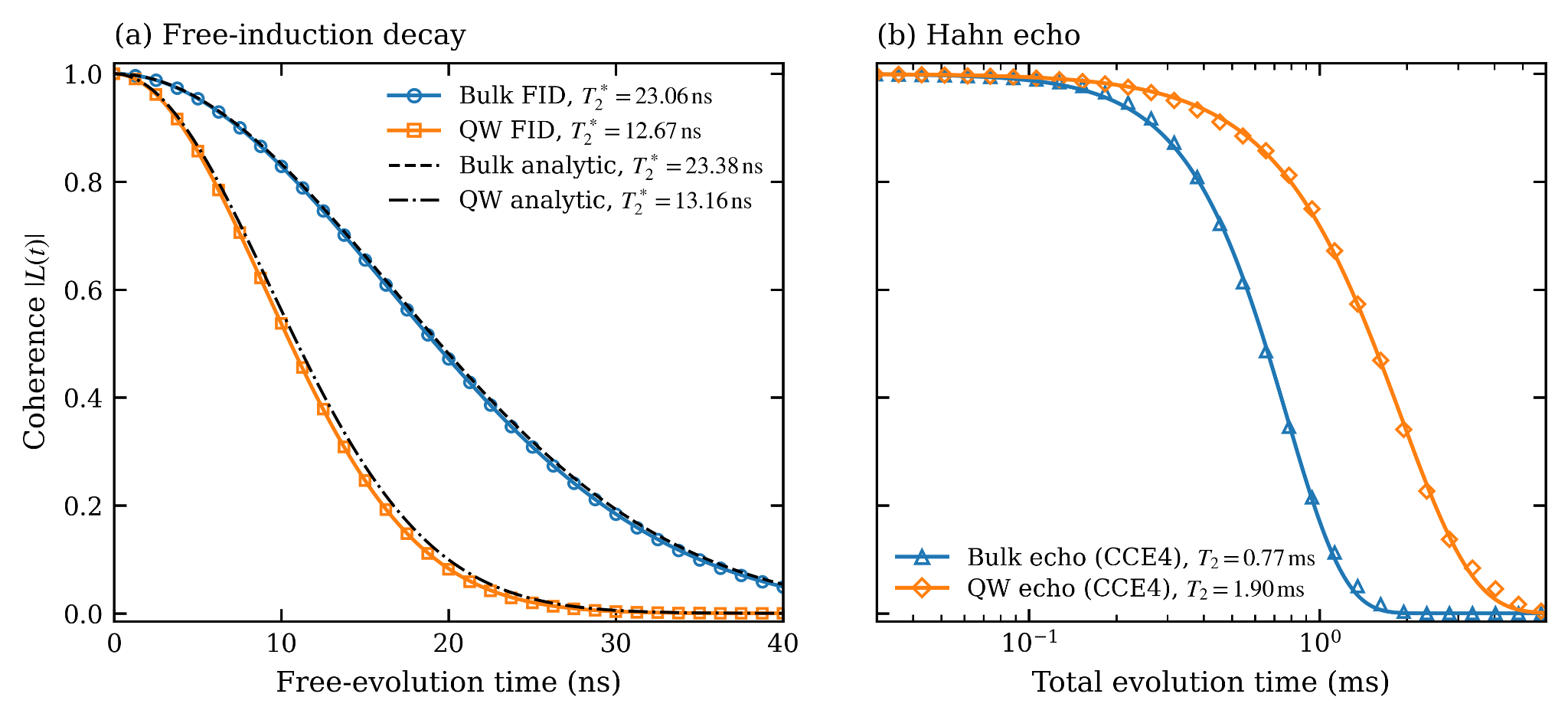}
    \caption{\textbf{Numerical coherence of the Cl-donor electron in bulk ZnSe and in the ZnSe/ZnMgSe quantum well.} \textbf{a,} Free-induction decay. Symbols show the CCE1 calculations, colored solid curves are Gaussian fits, and the black dashed and dash-dotted curves are the analytical predictions. \textbf{b,} Hahn-echo coherence. Symbols show the CCE4 calculations including the CCE2 cation correction, and solid curves are stretched-exponential fits. Blue and orange denote the bulk and quantum-well donors, respectively.}
    \label{fig:S5}
\end{figure}

Next, we extend our discussion to the ZnSe/Zn(Mg)Se heterostructure. Two effects of the Mg-containing barriers must be distinguished. First, the conduction-band offset confines the donor electron along the growth direction and changes the inverse participation of the envelope. Second, Mg replaces a fraction of the Zn nuclei in the barrier and introduces the spinful isotope $^{25}\mathrm{Mg}$.

Following the standard product construction used for shallow donors in semiconductor heterostructures \cite{betancur1995simple,mikhailov2003shallow}, we take the following variational ansatz in a cylindrical coordinate system:
\begin{equation}
    F_\lambda(\rho,z)=C_\lambda f_0(z)e^{-r/\lambda},
    \qquad
    r=\sqrt{\rho^2+z^2},
\end{equation}
where $\lambda$ is the variational donor-localization length and $C_\lambda$ normalizes the full three-dimensional envelope. For confinement along $z$, we use the square-well potential
\begin{equation}
    V_{\mathrm{QW}}(z)=
    \begin{cases}
        0, & |z|\leq L/2,\\
        V_c, & |z|>L/2,
    \end{cases}
\end{equation}
where $V_c$ is the barrier depth and $L$ is the well thickness. Solving for this potential, the normalized even-parity ground state is
\begin{equation}
    f_0(z)=\frac{1}{\sqrt{N_z}}
    \begin{cases}
        \cos(kz), & |z|\leq L/2,\\
        \cos(kL/2)e^{-\kappa(|z|-L/2)}, & |z|>L/2,
    \end{cases}
\end{equation}
with
\begin{equation}
    k=\frac{\sqrt{2m^*E_z}}{\hbar},
    \qquad
    \kappa=\frac{\sqrt{2m^*(V_c-E_z)}}{\hbar}.
\end{equation}
Continuity of the wavefunction and its derivative gives
\begin{equation}
    k\tan\!\left(\frac{kL}{2}\right)=\kappa,
    \qquad
    k^2+\kappa^2=\frac{2m^*V_c}{\hbar^2},
\end{equation}
with normalization length
\begin{equation}
    N_z
    =\frac{L}{2}
    +\frac{\sin(kL)}{2k}
    +\frac{\cos^2(kL/2)}{\kappa}.
\end{equation}

By solving the finite well with barrier $V_c$ for the ground state, all variables can be determined. The normalization constant is determined from
\begin{equation}
    1
    =2\pi C_\lambda^2
    \int_{-\infty}^{\infty}dz\,|f_0(z)|^2
    \int_0^\infty\rho\,d\rho\,
    e^{-2\sqrt{\rho^2+z^2}/\lambda}.
\end{equation}
Therefore, the total variational energy can be written as
\begin{equation}
    E_{\mathrm{var}}(\lambda)
    =E_z
    +\frac{\hbar^2}{2m^*\lambda^2}
    -\frac{e^2}{4\pi\epsilon_0\epsilon}
    \pi\lambda C_\lambda^2
    \int_{-\infty}^{\infty}dz\,|f_0(z)|^2e^{-2|z|/\lambda}.
\end{equation}
Minimization of this energy with respect to $\lambda$ gives the variational wavefunction of the electron. After this calculation, we calculate the inverse participation volume, which in the current case contains separate contributions from inside and outside the quantum well:
\begin{equation}
    J_{\mathrm{QW}}
    =\int|F_\lambda(\bm{r})|^4\,d^3r
    =J_w+J_b,
\end{equation}
where
\begin{equation}
    J_w
    =2\pi C_\lambda^4
    \int_{-L/2}^{L/2}dz\,|f_0(z)|^4e^{-4|z|/\lambda}
    \left(\frac{\lambda|z|}{4}+\frac{\lambda^2}{16}\right),
\end{equation}
and
\begin{equation}
    J_b
    =2\pi C_\lambda^4
    \int_{|z|>L/2}dz\,|f_0(z)|^4e^{-4|z|/\lambda}
    \left(\frac{\lambda|z|}{4}+\frac{\lambda^2}{16}\right).
\end{equation}
Define
\begin{equation}
    C_j
    =y_j\frac{I_j(I_j+1)}{3}
    \left(A_j^{\mathrm{HF}}\right)^2.
\end{equation}
Therefore, we can write the total Overhauser variance as
\begin{equation}
    \sigma_h^2
    =v_0\left\{
        C_{\mathrm{Se}}(J_w+J_b)
        +C_{\mathrm{Zn}}\left[J_w+(1-x)J_b\right]
        +xC_{\mathrm{Mg}}J_b
    \right\}.
\end{equation}

We estimate the parameters as follows. The Cl donor is taken to lie at the center of a ZnSe quantum well of width $L=3.4$~nm, surrounded by barriers with $x=0.137$. Based on the linear model and the value used in Ref.~\cite{von2024shadow}, we extract a well depth of $134.1$~meV. Minimization of the variational energy gives a variational length of $\lambda=2.8$~nm and a model binding energy $E_z-E_{\mathrm{var}}=47.9$~meV. The total inverse participation is $J_{\mathrm{QW}}=4.7\times10^{-3}$~nm$^{-3}$, where the barrier-to-total ratio is approximately $J_b/J=2.0\%$. Therefore, the barrier contribution is small.

Unlike for Zn and Se, we could not find reliable reported values for $A_{\mathrm{Mg}}^{\mathrm{HF}}$. The nuclear gyromagnetic ratio of Se is approximately three times larger than those of both Zn and Mg. In addition, Zn and Mg occupy similar lattice sites, the abundance of $^{25}\mathrm{Mg}$ is smaller, and the barrier makes only a small contribution to the total inverse participation. We therefore neglect the effect of Mg for the remainder of the analysis. With this assumption, we obtain
\begin{equation}
    T_{2,\rm G,QW}^*=\frac{\sqrt{2}\hbar}{\sigma_h}=13.2~\mathrm{ns}.
\end{equation}
The experimentally extracted value of 21.9 ns lies between the bulk and quantum-well predictions, As the confinement strength may vary based on the actual experimental condition.

\subsection{Simulation through PyCCE}

To verify the analytical model and calculations, we calculated the donor-electron coherence using the conventional cluster-correlation expansion implemented in PyCCE (Onizhuk and Galli, 2021) \cite{onizhuk2021pycce}. A zincblende ZnSe supercell was generated with the Cl donor on a Se site and the bound electron treated as a central spin $S=1/2$. The same envelope functions, isotope abundances, and site-dependent contact couplings defined in the analytical section were used. Nuclear spins within a bath radius of $5.0$~nm of the donor were retained. In the quantum-well barriers, the small $^{25}\mathrm{Mg}$ contribution was represented by an effective Zn-like cation coupling; because only approximately $2\%$ of the inverse participation lies in the barriers, this produces only a minor correction.

For the free-induction decay in Fig.~\ref{fig:S5}(a), only the Fermi-contact interaction was included. In this quasi-static limit, individual nuclei contribute independent phase shifts, so first-order CCE is sufficient and no cluster-connectivity cutoff is required. Fitting the numerical coherence gives $T^*_{2,\mathrm{CCE}}=23.06$~ns for the bulk donor and $T_{2,\mathrm{CCE,QW}}^*=12.67$~ns for the quantum-well donor, matching the analytical estimates. The residual difference reflects the finite discrete isotope realization used in the simulation, whereas the analytical calculation averages continuously over isotope occupation. This agreement verifies that the shorter quantum-well FID is captured by the increased inverse participation of the confined donor envelope.

For the Hahn echo in Fig.~\ref{fig:S5}(b), a single ideal refocusing pulse was applied. The contact interaction was supplemented by the electron--nuclear point-dipole interaction outside one localization length, and nuclear clusters were connected using a cluster-connectivity cutoff of $1.4$~nm. The dominant $^{77}\mathrm{Se}$ bath was evaluated through fourth-order CCE, while the weaker cation bath was retained through second order. The total coherence was taken as the product of these two contributions. The resulting curves were fitted to
\begin{equation}
    L_{\mathrm{echo}}(t)=\exp\!\left[-\left(\frac{t}{T_2}\right)^n\right]
\end{equation}
and yield $T_2=0.770$~ms with $n=2.17$ in bulk and $T_2=1.90$~ms with $n=1.72$ in the quantum well. We quote these CCE4 values as the highest-order numerical estimates of the host-nuclear-bath contribution.

Although confinement shortens $T_2^*$, the calculated Hahn-echo time is longer in the quantum well. We attribute the longer echo to the stronger spatial variation of the confined envelope function, which spreads the hyperfine couplings of neighboring nuclei and detunes their nuclear--nuclear flip-flop transitions. A larger spread in on-site hyperfine energies raises the energy cost of a pairwise flip-flop and suppresses a fraction of the bath dynamics that the echo cannot refocus.

The millisecond values isolate the host-nuclear spectral-diffusion channel and do not imply that the experimental coherence must reach the same scale. Residual dephasing persists even after isotope purification, as resident electrons in an isotopically purified ZnSe/(Zn,Mg)Se quantum well exhibited a low-field dephasing time of about $7$~ns, attributed to spin--orbit-mediated relaxation \cite{kopteva2019spin}, while isotopically purified ZnSe zinc-vacancy ensembles reached approximately $0.4~\mu$s at low temperature and were limited by anisotropic exchange \cite{kirstein2021extended}. Spin--orbit and phonon processes, donor--donor interactions, charge fluctuations, tunneling or ionization, and pulse imperfections may therefore limit the measured coherence before the nuclear-bath value is reached.

\clearpage

\section{Ultrafast optical spin rotation}

\subsection{Model of the spin rotation}

We use a red-detuned ultrafast pulse to coherently rotate the donor electron spin, following established approaches to ultrafast optical spin control in semiconductor spin systems \cite{clark2007quantum,kliX,83Xt}. We denote the two electron-spin ground states by $\ket{0}$ and $\ket{1}$ and the two donor-bound-exciton states by $\ket{2}$ and $\ket{3}$. The broadband pulse couples both ground states to the excited-state manifold, with time-dependent Rabi frequencies
\begin{equation}
    \Omega_{mi}(t)
    =
    \frac{
        \boldsymbol{\mu}_{mi}\cdot\mathbf{E}(t)
    }{\hbar},
    \qquad
    i\in\{0,1\},
    \quad
    m\in\{2,3\}.
\end{equation}

The ultrafast laser is red-detuned from the manifold by $700$~GHz. When the optical detunings are large compared with the coupling and decay rates, the excited states are only virtually populated. The effective ground-state Hamiltonian is
\begin{align}
    H_{\mathrm{eff}}(t)
    ={}&
    -\frac{\hbar}{4}
    \sum_{\substack{i=0,1\\m=2,3}}
    \frac{
        \left|\Omega_{mi}(t)\right|^2
    }{\Delta_{mi}}
    \ket{i}\!\bra{i}
    \nonumber\\
    &+
    \frac{\hbar}{2}
    \left[
        \Omega_{\mathrm{eff}}(t)
        \ket{0}\!\bra{1}
        +\mathrm{H.c.}
    \right],
    \label{eq:effective_spin_H}
\end{align}
where $\Delta_{mi}$ is the detuning of the pulse from the $\ket{i}\leftrightarrow\ket{m}$ transition. The diagonal terms describe the optical AC Stark shifts. The off-diagonal stimulated-Raman coupling is
\begin{equation}
    \Omega_{\mathrm{eff}}(t)
    =
    -\frac{1}{4}
    \sum_{m=2,3}
    \Omega_{m0}^*(t)\Omega_{m1}(t)
    \left(
        \frac{1}{\Delta_{m0}}
        +
        \frac{1}{\Delta_{m1}}
    \right).
    \label{eq:effective_Raman}
\end{equation}
The relative amplitudes and phases of the four optical couplings determine the spin-rotation axis.

For fixed pulse duration, repetition rate, spatial mode, and polarization,
\begin{equation}
    \left|\Omega_{mi}(t)\right|^2\propto P,
\end{equation}
where $P$ is the average pulsed-laser power at the sample. It follows from Eq.~\eqref{eq:effective_Raman} that
\begin{equation}
    \Omega_{\mathrm{eff}}(t)\propto P.
\end{equation}
The rotation angle produced by one pulse is
\begin{equation}
    \theta(P)
    =
    \int_{-\infty}^{+\infty}
    \left|\Omega_{\mathrm{eff}}(t)\right|\,dt,
    \label{eq:rotation_angle}
\end{equation}
and therefore increases linearly with average pulse power under fixed experimental conditions.

The power-dependent single-pulse spin-rotation is fitted as
\begin{equation}
    I(P)
    =
    I_{\mathrm{0}}
    +
    C_0 e^{-P/P_d}
    \cos\!\left(\alpha P+\phi\right),
    \label{eq:Rabi_power_fit}
\end{equation}
where $\alpha$ converts average pulse power to rotation angle, $\phi$ is a phase offset, $P_d$ characterizes the loss of oscillation contrast, and $I_{\mathrm{0}}$ and $C_0$ are the intensity offset and amplitude. The exponential envelope captures the observed power-dependent loss of contrast, including excitation-induced decoherence.

\subsection{Details of the pulsed-laser sequence}

The resonant excitation is supplied by a TOPTICA continuous-wave laser, modulated with a Brimrose acousto-optic modulator and prepared in the $H$ polarization state. Ultrafast pulses from a $76$~MHz mode-locked laser are selected with a Conoptics electro-optic pulse picker, reducing the repetition rate to $1.52$~MHz. The pulse-picker extinction ratio is approximately $80{:}1$. After converting the laser pulses to $\sim436$~nm through the second-harmonic generation using a BBO-crystal, the extinction ratio is approximately $10^4{:}1$. The pulsed laser is linearly polarized at $45^\circ$ relative to both $H$ and $V$; its focused spot size is approximately $1~\mu$m.

\subsection{Magnetic-field dependence of the spin precession period}

\begin{figure}[H]
    \renewcommand{\figurename}{FIG.}
    \renewcommand{\thefigure}{S\arabic{figure}}
    \setcounter{figure}{5}
    \centering
    \includegraphics[width=200 pt]{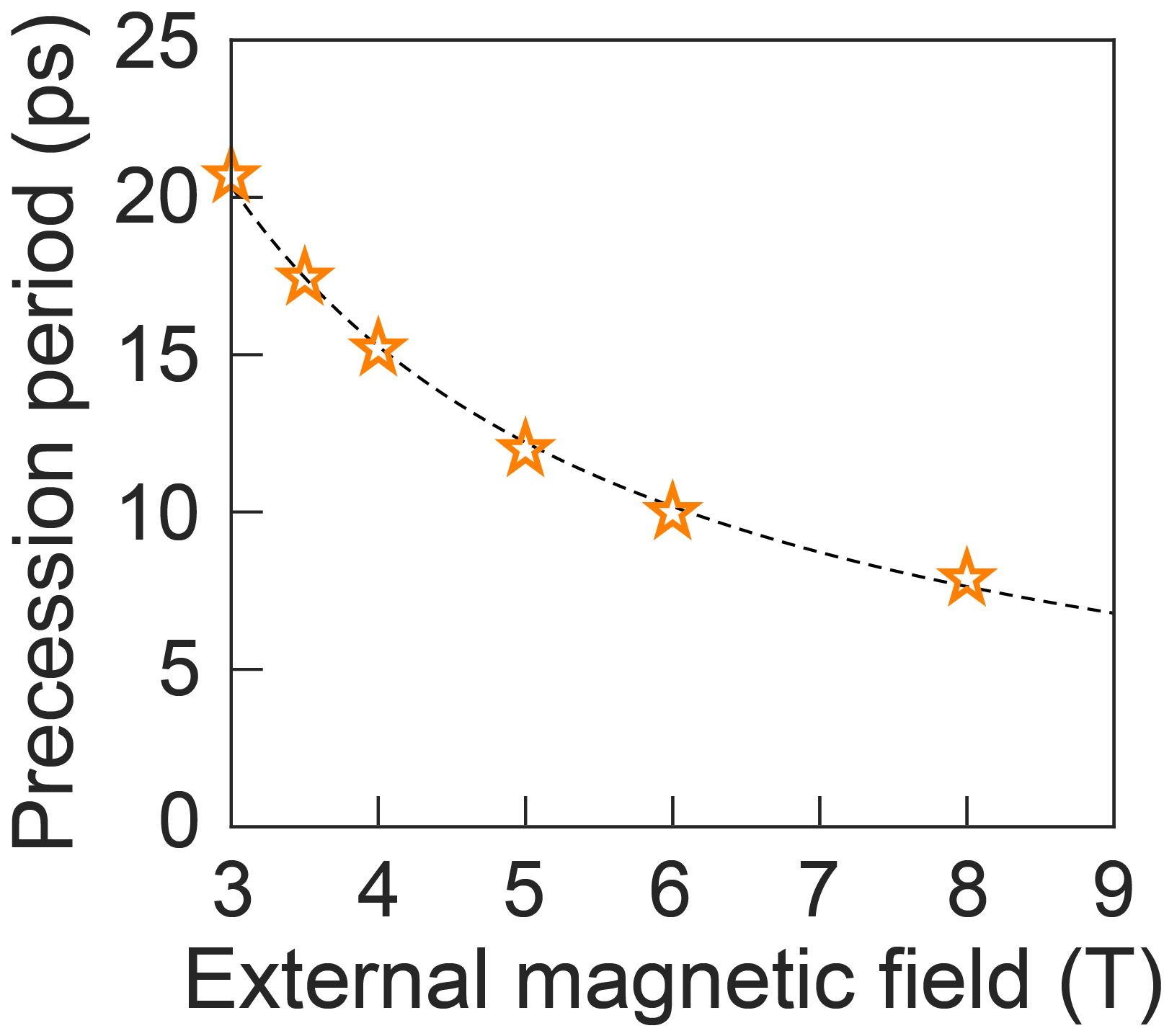}
    \caption{\textbf{Magnetic-field dependence of the spin precession period.}
    Markers show the precession periods extracted from the Ramsey 
    measurements at different external magnetic fields. The black dashed 
    line shows the calculated spin Larmor precession period, using
    $T_{\mathrm{p}}=h/(g_e\mu_B B)$ and $g_e=1.17$.}
    \label{fig:S6}
\end{figure}

To verify the origin of the oscillations observed in the Ramsey measurements, 
we extract the oscillation period at different external magnetic fields. 
For an electron spin in a magnetic field of $B$, the Larmor precession frequency is
\begin{equation}
    \omega_L = \frac{g_e \mu_B B}{\hbar},
\end{equation}
corresponding to a precession period
\begin{equation}
    T_{\mathrm{p}} = \frac{2\pi}{\omega_L}.
\end{equation}
As shown in Fig.~\ref{fig:S6}, the measured precession period decreases with 
increasing magnetic field. The dashed curve is calculated using $g_e=1.17$ extracted from the photoluminescence spectrum, which is consistent with the measurement interference results.

\clearpage

\bibliographystyle{apsrev4-2}
\bibliography{refs}